\documentclass[rsi,reprint,amsmath,amssymb,author-numerical]{revtex4-1} 

\usepackage{graphicx}
\usepackage{dcolumn}
\usepackage{bm}
\usepackage[version=4]{mhchem}

\usepackage[utf8]{inputenc}
\usepackage[T1]{fontenc}
\usepackage{mathptmx}
\usepackage{etoolbox}
\usepackage{booktabs}
\usepackage{array}
\usepackage{float}
\usepackage{ragged2e}
\usepackage{xcolor}
\usepackage[final]{changes} 

\makeatletter
\def\@email#1#2{%
 \endgroup
 \patchcmd{\titleblock@produce}
  {\frontmatter@RRAPformat}
  {\frontmatter@RRAPformat{\produce@RRAP{*#1\href{mailto:#2}{#2}}}\frontmatter@RRAPformat}
  {}{}
}%
\makeatother
\begin{document}


\title[Sample title]{Development of a Modular Optically Detected Magnetic Resonance Setup\\for Optical Experiments in a Variable Temperature Insert}
\author{Anh Tong$^{1,2,4,a}$, Andreas Bauer$^{1,2}$, Markus Kleinhans$^{1}$, James S. Schilling$^{1}$, Christian H. Back$^{1,2,4}$, Karl D. Briegel$^{4,5}$, Fabian A. Freire-Moschovitis$^{4,5}$, Dominik B. Bucher$^{4,5}$, Christian Pfleiderer$^{1,2,3,4}$}

\affiliation{$^{1}$Technical University of Munich, TUM School of Natural Sciences, Physik Department, James-Franck-Str. 1, 85748 Garching, Germany}
\affiliation{$^{2}$Centre for QuantumEngineering (ZQE), Technical University of Munich, Am Coulombwall 3a, 85748 Garching, Germany}
\affiliation{$^{3}$Heinz Maier-Leibnitz Zentrum, Technical University of Munich, Lichtenbergstraße 1, 85748 Garching, Germany}
\affiliation{$^{4}$Munich Center for Quantum Science and Technology (MCQST), Schellingstraße 4, 80799 München, Germany}
\affiliation{$^{5}$Technical University of Munich, TUM School of Natural Sciences, Department of Chemistry, Lichtenbergstrasse 4, 85748 Garching, Germany}

\date{\today}

\begin{abstract}
We developed an optically detected magnetic resonance (ODMR) setup designed for compatibility with a widely used, commercially available helium bath cryostat equipped with a variable temperature insert. The optical path extends nearly two meters, spanning the full length of the cryostat insert, enabling excitation of the nitrogen-vacancy (NV) centers and detection of the resulting fluorescence from outside the cryostat. The setup preserves optical alignment and beam quality along this extended path, allowing integration into existing cryogenic systems without significant modifications. We demonstrate the setup’s performance by measuring the temperature dependence of the resonance signal and its behavior under small applied magnetic fields, as well as the magnetic transition of a SrRuO$_3$ sample, thereby showcasing the feasibility of NV magnetometry on a sample in constrained cryogenic environments.

\end{abstract}

\maketitle

\noindent\textit{a) Author to whom correspondence should be addressed: anh.tong@tum.de}

\section{\label{sec:level1} Introduction}
The nitrogen–vacancy (NV) center in diamond provides a direct means to detect magnetic fields through its spin-dependent energy levels, which shift in response to the local magnetic field via the Zeeman effect. By combining optical laser excitation with microwave driving, the NV electronic spin state can be initialized, manipulated, and read out optically \cite{Jensen2017, Rondin2014, Xu2023}. This technique enables quantitative magnetic field measurements with high sensitivity and sub-micrometer spatial resolution. Owing to these capabilities, NV magnetometry has become a promising tool for probing complex magnetic textures in condensed matter systems\,\citep{gross2018realspace, du2017control, Luethi2025}. Micro- and nanoscale spin configurations such as domain walls, skyrmions, and helical textures arise from competing interactions and can often play a key role when tuning electronic and magnetic properties near phase boundaries. For instance, NV magnetometry has enabled real-space reconstruction of the stray field and full spin structure of a single Néel-type skyrmion in a magnetic multilayer, revealing depth-dependent features that are inaccessible to most other probes\,\citep{Dovzhenko2018}.

Extending such measurements to quantum materials requires extreme conditions such as low temperatures, high magnetic fields, and often high pressures, however, most NV magnetometry setups are tailored to room-temperature operation or specialized optical cryostats. A practical and versatile strategy is, therefore, to integrate NV-based measurements into commercially available cryogenic infrastructures such as helium bath cryostats with variable temperature inserts (VTIs) or diamond anvil high pressure cells (DACs)\,\citep{jorba_high-pressure_2022} by adapting the NV-experiments' optics to the cryogenic infrastructure constraints rather than adapting the cryogenic systems themselves.

Here, we present a modular ODMR setup developed for an Oxford Instruments recondensing IntegraAC cryostat equipped with a 16\,T superconducting magnet and a VTI \replaced{capable of operation between 1.6\,K and 300\,K}{covering a temperature range from 1.6\,K to 300\,K}. This design enables NV magnetometry in a standard cryogenic environment while maintaining high optical stability and measurement reproducibility. Moreover, the setup is fully compatible with diamond anvil pressure cells, allowing future studies of correlated materials under combined extreme conditions\,\citep{Hsieh2019, Lesik2019, Yip_2019}.

This work is structured as follows. Section\,\ref{ExperimentalSetup} provides an overview of the technique and its challenges (A), describes the design of the ODMR setup including the optical head, sample stick, and support structures (B), and details the assembly, pre-alignment procedures, cryostat installation (C), and measurement electronics (D). Section \ref{DataAnalysis} explains the analysis and evaluation of the recorded spectra. Section \ref{T_Dep} discusses temperature-dependent measurements, and Section \ref{Magn_Dep} field-dependent measurements. Section \ref{SampleSRO} presents measurements on a SrRuO$_3$ sample comparing it with standard MPMS measurements. \added{Section \ref{SensitivityEstimate} provides an order-of-magnitude estimate of the sensitivity of our NV-ensemble ODMR setup in comparison to conventional bulk magnetometry techniques such as VSM.} The paper concludes with Section\,\ref{Discussion_Concl} summarizing the main findings.

\section{Experimental setup} \label{ExperimentalSetup}

\subsection{\label{sec:meth1}Overview}
Optically detected magnetic resonance forms the basis of NV-center magnetometry. The negatively charged NV center (NV$^-$) features a spin-triplet ground state ($^3A_2$) and an optically accessible spin-triplet excited state ($^3E $). Under green laser excitation, the electronic spin of the NV center is polarized into the bright $m_s = 0$ ground state. When a microwave (MW) field resonant with the $m_s = 0 \leftrightarrow \pm1$ transitions is applied, population transfers into the darker $m_s = \pm1$ states, reducing fluorescence. Sweeping the microwave frequency reveals resonances whose Zeeman splitting directly reflects the local magnetic field experienced by the NV center under an applied field. This technique provides a quantitative means of detecting local magnetic fields with high sensitivity and potential sub-micrometer spatial resolution.

Implementing ODMR in high-field helium bath cryostats with variable temperature inserts, such as the Oxford Instruments recondensing IntegraAC, is technically demanding. These cryostats, optimized for transport or thermodynamic measurements, provide limited optical access, long optical paths of up to two meters and in our case only a 30\,mm VTI bore for all in-situ components. Precise optical alignment and mechanical stability are therefore critical, as small misalignments of the optical path strongly affect signal strength and reproducibility.

To meet these constraints, we developed a modular setup that adapts the optical experiment to a standard commercial cryostat without requiring modifications to the cryostat itself. Our setup comprises three main components: (1) an external optical head for laser excitation and fluorescence detection, (2) a custom sample stick with integrated microwave and optical components, and (3) a rail-guided platform mounted above the cryostat that ensures reproducible positioning of the optical head with respect to the sample after reinsertion. A modular frame built from standard extrusion profiles (e.g. ITEM) can support the sample stick and optical head outside the cryostat at room temperature and enables pre-alignment and testing before the installation into the cryostat. Our setup is designed to be compatible with diamond anvil high pressure cells, extending its applicability to combined high-field, low-temperature, and high-pressure experiments.

Sections \ref{sec:meth2}, \ref{sec:meth4}, \ref{sec:meth5} detail the individual components and the typical measurement procedure.

\subsection{\label{sec:meth2}Design of the ODMR setup}
\subsubsection{Optical Head}
\begin{figure}[t!]
    \centering
    \includegraphics[width=0.5\textwidth]{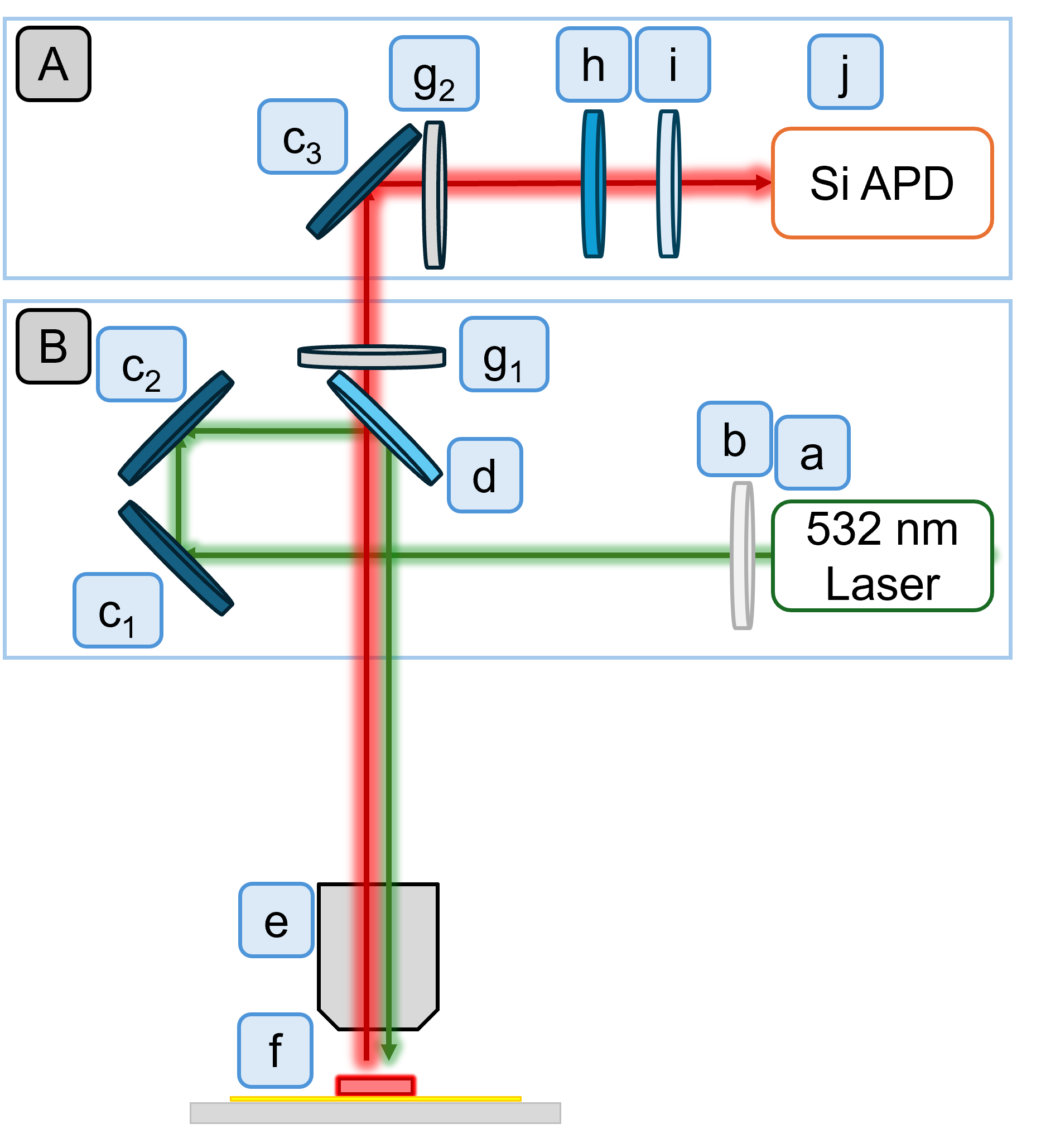}
    \vspace{0.2cm}

    \begin{tabular}{@{}l@{}l@{}}
    \toprule
    \parbox[t]{0.1\columnwidth}{\raggedright\textbf{Label}} & \parbox[t]{0.8\columnwidth}{\raggedright \textbf{Component}} \\
    \midrule
    \parbox[t]{0.1\columnwidth}{\raggedright a} & \parbox[t]{0.8\columnwidth}{\noindent\justifying Optical fiber ($\varnothing$ 50\,$\mu$m, 0.22\,NA, Low OH, FC/PC-FC/PC Fiber Patch Cable, Achromatic FiberPort, FC/PC, f=10.0\,mm, 400-700\,nm, $\varnothing$1.64 mm Waist)} \\
    \parbox[t]{0.1\columnwidth}{\raggedright b} & \parbox[t]{0.8\columnwidth}{\noindent\justifying Collimator} \\
    \parbox[t]{0.1\columnwidth}{\raggedright$c_1$--$c_3$} & \parbox[t]{0.8\columnwidth}{\noindent\justifying Plano mirrors (Protected Silver Mirror)} \\
    \parbox[t]{0.1\columnwidth}{\raggedright d} & \parbox[t]{0.8\columnwidth}{\noindent\justifying Longpass dichroic mirror (Longpass Dichroic Mirror, 567\,nm Cut-On)} \\
    \parbox[t]{0.1\columnwidth}{\raggedright e} & \parbox[t]{0.8\columnwidth}{\noindent\justifying Olympus LMPLFLN 10x objective, 
       long working distance 21\,mm, NA 0.25} \\
    \parbox[t]{0.1\columnwidth}{\raggedright f} & \parbox[t]{0.8\columnwidth}{\noindent\justifying Coplanar waveguide (CPW) with NV diamond chip on top} \\
    \parbox[t]{0.1\columnwidth}{\raggedright $g_1$--$g_2$} & \parbox[t]{0.8\columnwidth}{\noindent\justifying Longpass Filter, Cut-On Wavelength: 650\,nm} \\
    \parbox[t]{0.1\columnwidth}{\raggedright h} & \parbox[t]{0.8\columnwidth}{\noindent\justifying Plano-convex spherical lens (N-BK7 Plano-Convex Lens, $\varnothing$\,1", f=150\,mm, AR Coating: 400-1100\,nm)} \\
    \parbox[t]{0.1\columnwidth}{\raggedright i} & \parbox[t]{0.8\columnwidth}{\noindent\justifying Neutral density filter (Absorptive ND Filter, Optical Density: 0.7)} \\
    \parbox[t]{0.1\columnwidth}{\raggedright j} & \parbox[t]{0.8\columnwidth}{\noindent\justifying Si APD module (A-CUBE-S3000-10, Si-APD-Module, D=3\,mm, DC-10\,MHz)} \\
    \bottomrule
    \end{tabular}

    \caption{Schematic of the optical components and their arrangement. The components are labeled and listed in the legend above. The schematic shows the light path of the excitation laser as well as the path of the collected fluorescence.}
    \label{fig:SchematicOpticalComponents}
\end{figure}

All optical components located outside the cryostat, above the cryostat flange, are mounted on a two-level breadboard construction, separating the detection (A) and excitation (B) paths vertically as shown in Fig.\,\ref{fig:SchematicOpticalComponents}. The breadboards are enclosed in opaque covers to prevent stray light contamination and ensure signal integrity. This arrangement is referred to as the modular optical head.

The optical path is designed to guide and align 532\,nm laser excitation light onto the sample and to collect NV fluorescence with high spatial stability and minimal loss over the full optical access of the cryostat. For safety and modularity, the laser is fiber-coupled (a), and the optical head receives the light through a collimator (b), converting it into a free-space beam. \added{We employ a free-space optical configuration. A fiber-based approach at the sample position was considered not suitable due to space constraints in the sample region associated with the sample-stick design and its intended compatibility with future integration of a diamond anvil cell.}

Two mirrors mounted on kinematic gimbal mounts (c$_1$,c$_2$) immediately follow the collimator output. These provide two degrees of freedom each to align the beam vertically and horizontally with respect to the optical axis and the worktop plane, respectively. This configuration facilitates precise beam steering using the standard “beam walking” technique: the beam is adjusted to be parallel to the worktop plane by iteratively tuning the mirror angles to maintain beam position through fixed apertures.

The aligned beam is then directed onto a longpass dichroic mirror (d) that reflects wavelengths below 567\,nm, thus reflecting the 532\,nm excitation beam, and transmits longer wavelengths. This mirror is mounted in a fine-tunable holder to ensure angular stability and fine tuning of its reflection plane. The reflected beam passes vertically downward through a high-vacuum CF flange, which serves as the interface between the external optical components and the interior of the cryostat, into the cryostat.

Inside the cryostat, the beam is focused by an infinity-corrected, long working distance objective (e) onto the sample, for example, a bulk diamond chip used for initial testing, with dimensions of $3\times3\times0.5\,\mathrm{mm}^3$, a \{100\}$\pm3^\circ$ crystallographic surface orientation, and a typical NV density of about $4.5$\,ppm homogeneously distributed throughout the entire crystal, (f). The 532\,nm excitation induces fluorescence from the NV centers. For clarity, we note already at this point that the external magnetic field is applied perpendicular to the (100) diamond surface. The green excitation laser spot diameter at the sample is estimated using standard Gaussian beam optics, taking into account the underfilling of the objective aperture in our setup. For a beam diameter of 5\,mm and a focal length of 18\,mm, this yields a diffraction-limited spot diameter on the order of 5-6\,$\mu$m for 532\,nm wavelength.

The fluorescence is emitted in a $4\pi$ solid angle, but only a fraction (an objective with an ideal NA of 0.25 in air captures approximately 1.5\,$\%$ of isotropic emission) is collected by the same objective. The collected fluorescence passes upward through the viewport, is collimated, and guided back through the dichroic mirror at the top. It then passes through a longpass filter (g$_1$) with a cut-on wavelength of 650\,nm to suppress residual green excitation light. It is then reflected by a third mirror (c$_3$) by 90$^\circ$, passes through a second longpass filter (g$_2$) to further suppress any stray light, and is finally focused using a plano-convex lens (h).
This bundled fluorescence beam is attenuated with a neutral density (ND) filter (i), which is employed to prevent detector saturation given the high NV concentration and resulting strong emission from this type of sample, and directed into a silicon avalanche photodiode (j) for detection. In addition to protecting the detector, the combination of an ND filter with the Si-APD provides flexibility to accommodate varying fluorescence intensities. While the sample shown here exhibits strong emission, other investigated diamond samples (not shown) can be significantly dimmer. Interchangeable ND filters therefore enable controlled attenuation to keep the APD within its linear regime for bright samples, while allowing minimal attenuation for weaker signals, ensuring optimal signal-to-noise across different sample types.

\subsubsection{\label{sec:meth3}Sample Stick}
\begin{figure}[t!]
    \centering
    \includegraphics[width=1\linewidth]{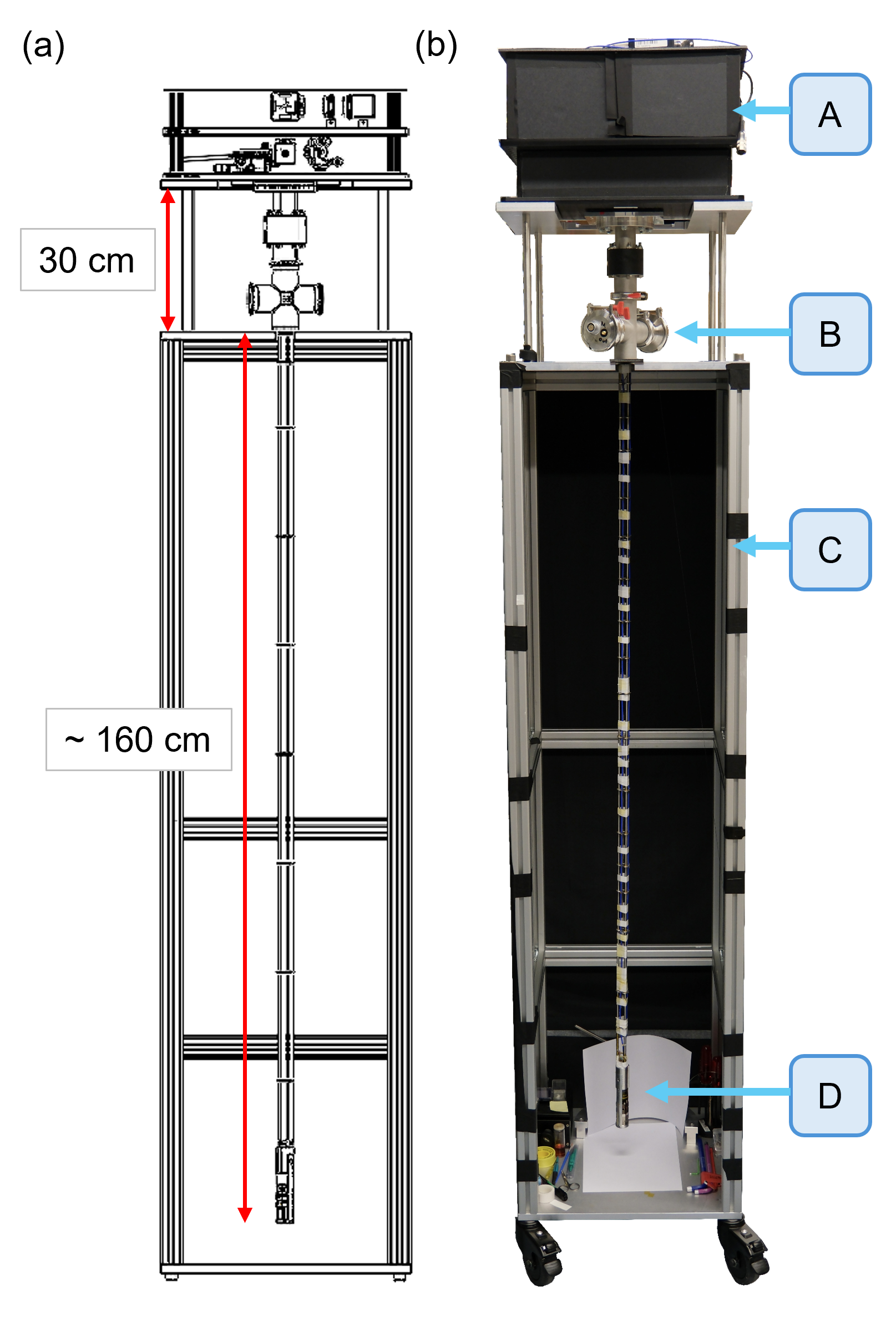}
    \caption{(a) CAD assembly of the sample stick and its support structure used for optical alignment at room temperature prior to installation in the cryostat. Key dimensions and features are indicated. (b) Photograph of the assembled sample stick with support structure for comparison. (A) Optical head and sample stick for the VTI (B) in its support frame for room-temperature storage and testing (C) and (D) shows the sample stick sample cage with a detailed view in the following figure.}
    \label{fig:SampleStickFrame}
\end{figure}

The sample stick, shown in Fig.\,\ref{fig:SampleStickFrame} together with its support frame, was custom-designed to meet the mechanical and spatial constraints imposed by the cryostat’s VTI, which limits the maximum outer diameter to 30\,mm. It consists of three ~160\,cm long, stainless steel rods arranged in a triangular configuration and interference-fitted (press-fit/nitrogen fit) into a retaining cylindrical block at the top that features an axial through-hole for optical access. This retaining block is press-fit as well into a KF40 cross flange, providing both mechanical stability and a modular interface to the cryostat. The total length of the light path is approximately 190\,cm, measured from the mounting point of the optical head at the top to the sample position housed in the sample holder at the lower end. 

\begin{figure}[t!]
    \centering
    \includegraphics[width=1\linewidth]{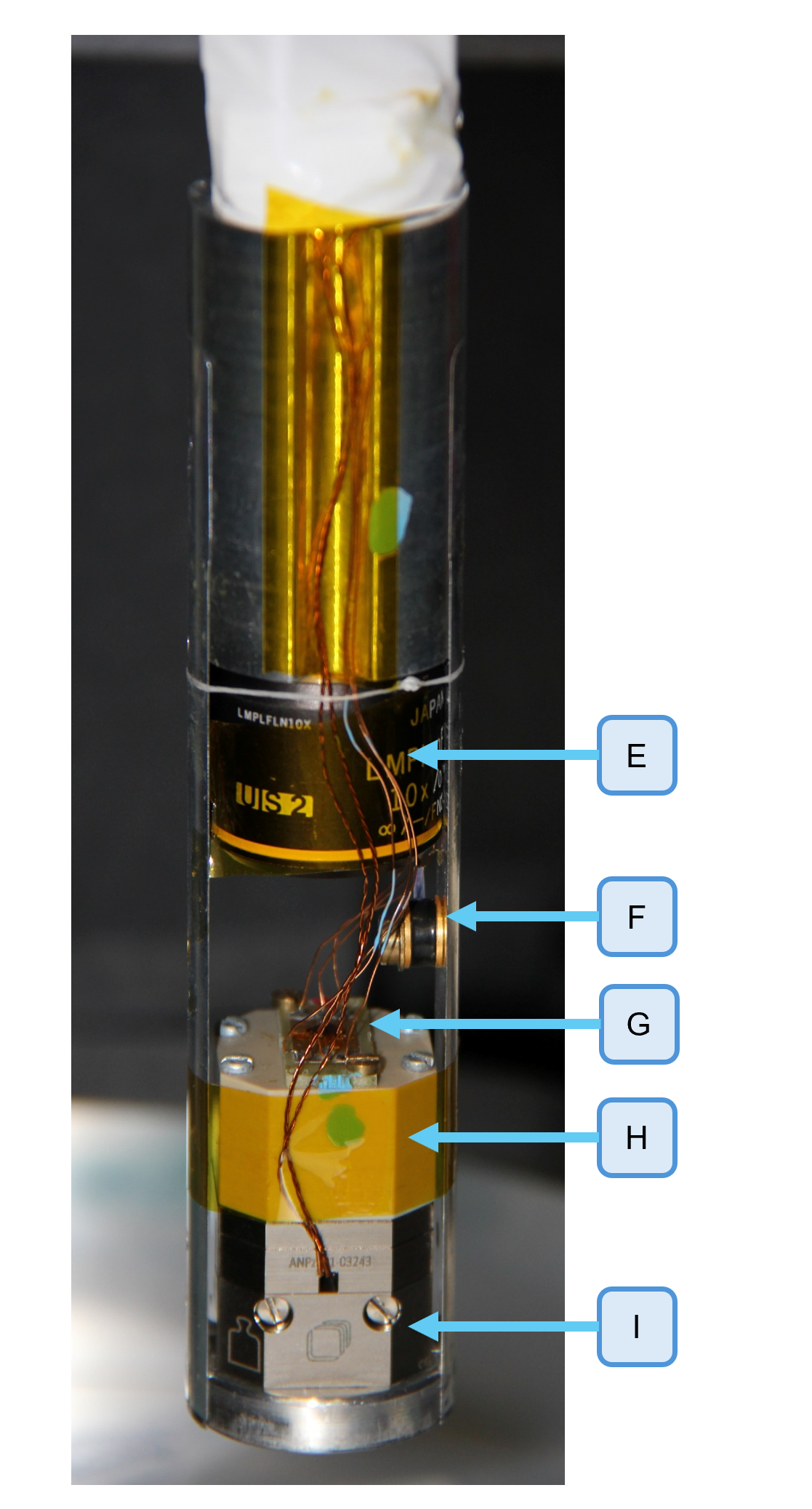} 
  
    \begin{tabular}{@{}l@{}l@{}l@{}}
    \toprule   
    \parbox[t]{0.1\columnwidth}{\raggedright\textbf{Label}} & \parbox[t]{0.4\columnwidth}{\raggedright \textbf{Component}} & \parbox[t]{0.5\columnwidth}{\raggedright \textbf{Model/Specification}}\\
    \midrule
    \parbox[t]{0.1\columnwidth}{\raggedright E} & \parbox[t]{0.4\columnwidth}{\noindent\justifying Objective} & \parbox[t]{0.5\columnwidth}{\noindent\justifying Olympus LMPLFLN 10x, working distance 21\,mm, NA 0.25} \\
    \parbox[t]{0.1\columnwidth}{\raggedright F} & \parbox[t]{0.4\columnwidth}{\noindent\justifying Temperature sensor} & \parbox[t]{0.5\columnwidth}{\noindent\justifying Lakeshore Cernox calibrated} \\
    \parbox[t]{0.1\columnwidth}{\raggedright G} & \parbox[t]{0.4\columnwidth}{\noindent\justifying CPW} & \parbox[t]{0.5\columnwidth}{\noindent\justifying Custom} \\
    \parbox[t]{0.1\columnwidth}{\raggedright } & \parbox[t]{0.4\columnwidth}{\noindent\justifying Diamond chip} & \parbox[t]{0.5\columnwidth}{\noindent\justifying $3\times3\times0.5\,\mathrm{mm}^3$, NV concentration ~\,4.5\,ppm homogeneously distributed throughout bulk} \\
    \parbox[t]{0.1\columnwidth}{\raggedright H} & \parbox[t]{0.4\columnwidth}{\noindent\justifying Spacer} & \parbox[t]{0.5\columnwidth}{\noindent\justifying PEEK material, custom} \\
    \parbox[t]{0.1\columnwidth}{\raggedright I} & \parbox[t]{0.4\columnwidth}{\noindent\justifying Linear \textit{z}-piezo actuator} & \parbox[t]{0.5\columnwidth}{\noindent\justifying Attocube ANPz101/RES/HL/LT/UHV} \\
    \bottomrule
    \end{tabular}  

    \caption{Detailed view of the sample holder at the lower end of the sample stick, showing the objective (E), the temperature sensor (F), a diamond chip mounted on a coplanar waveguide (G), the sample holder (H), and the piezo actuator at the base (I). Specific details may be found in the table above.}
    \label{fig:SampleStickFrame_detail}
\end{figure}

A detailed view of the sample holder is shown in Fig.\,\ref{fig:SampleStickFrame_detail}. The rigid aluminum holder at the bottom end holds all key components in a compact, mechanically stable configuration: the objective lens (E), a Cernox temperature sensor (F), the custom coplanar waveguide for microwave delivery, and the sample itself (G) on a spacer (H), as well as the non-magnetic linear z-piezo actuator (I) for fine axial positioning. This tightly integrated assembly ensures robust alignment within the narrow VTI bore and supports reproducible operation across thermal cycles. Additionally, equidistantly spaced baffles are installed along the sample stick to reduce radiative heat transfer and suppress stray light propagation.  

Optical access is provided through a high-vacuum CF flange fitted with a wedged 1.5" UVFS window (AR-coated, 350–700\,nm), enabling stable imaging and excitation through the full cryostat length. The microscope objective is mounted just above the sample and remains fixed relative to the sample holder while the piezo actuator can move the sample vertically to adjust the focus. Microwave delivery is realized via a coplanar waveguide with solder pads and a 90$^{\circ}$ miniature coaxial RF connector (MMCX), connected to vacuum-compatible SMA feedthroughs mounted at the top of the sample stick on the sides of the KF40 top flange. Thermometer, piezo actuator, and other electrical lines are routed along the sample stick and exit through 12-pin circular push-pull connectors (LEMO-type) at the top flange, separated into excitation and detection sides to reduce cross-talk. The design ensures mechanical stability, thermal compatibility, and reliable operation across multiple cool-down cycles.

\subsubsection{\label{sec:meth3.5}Support Table and Frame for Pre-Alignment and Handling}
To enable reproducible optical alignment and safe handling of the setup components, we constructed two modular aluminum support frames built from standard extrusion profiles: a frame for pre-aligning and testing at room temperature, and a cryostat-mounted platform system. The pre-alignment frame (Fig.\,\ref{fig:SampleStickFrame}\,C) holds the sample stick (Fig.\,\ref{fig:SampleStickFrame}\,B) vertically and provides mechanical support for the optical head (Fig.\,\ref{fig:SampleStickFrame}\,A), allowing full access to both the optical components at the top and the sample region at the bottom (Fig.\,\ref{fig:SampleStickFrame}\,D). The enclosure minimizes stray light and facilitates testing and alignment procedures prior to installation into the cryostat. 

\begin{figure}[t!]
    \centering
    \includegraphics[width=1\linewidth]{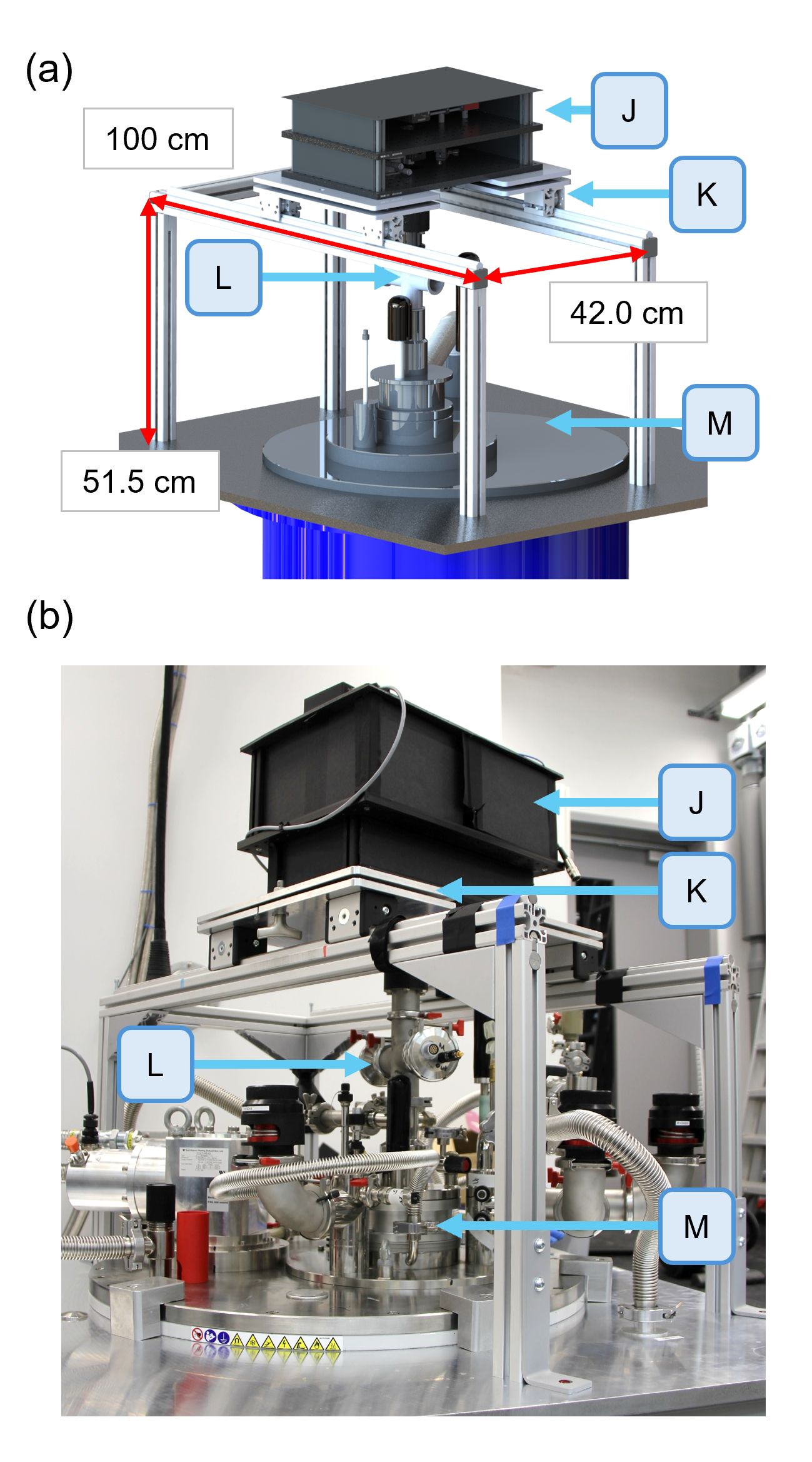}
    \caption{(a) CAD model of the modular aluminum support table with a rail-guided platform (K), designed for reproducible positioning of the optical head (J) above the sample stick (L) mounted in the cryostat (M). The sliding platform ensures precise and repeatable alignment. Key structural features and dimensions are annotated. (b) Photograph of the assembled support table with the optical head mounted on the sliding platform for comparison.}
    \label{fig:CryoTable}
\end{figure}

On top of the cryostat, see Fig.\,\ref{fig:CryoTable}\,(M), a height-adjusted table supports a rail-guided platform (K) for the optical head (J). The platform consists of two stacked levels, with spring elements in-between enabling fine vertical and angular adjustment. The height is tuned via nuts and bolts that compress the springs and securely hold the optical head in place. This design allows for safe, low-stress installation of the optical head and ensures reproducible alignment.

\subsection{\label{sec:meth4}Setup Assembly, Pre-Alignment, and Cryostat Installation}
To ensure consistent and reproducible alignment after each cooling cycle, we follow a well-defined procedure for mounting and aligning the optical system. The sample stick is first inserted into the cryostat. Then the rail-guided platform is mounted on the support table on top of the cryostat described above and is positioned horizontally above it. The optical head is loosely attached to the top of the sample stick, and the upper level of the platform is then gently lowered by tightening adjustment nuts, compressing spring elements placed between the two stacked platform levels described before and securely seating the optical head. 

This setup allows for fine vertical and angular adjustment and mechanically decouples stress from the stick. After securing the optical head, electrical connections are made via dedicated plugs and sockets on the sides of the KF40 cross of the sample stick. Typically, only minimal optical realignment is required, most notably a fine angular correction of the dichroic mirror to re-optimize fluorescence collection. In rare cases, only a small angular adjustment of less than 1$^\circ$ was necessary to reach the maximum signal intensity, while in most re-installations or after cooldown no adjustment was required. This modular mounting approach ensures high reproducibility and mechanical stability over multiple cooldown and reassembly cycles. The platform also enables the optical head to be temporarily moved sideways to access the cryostat’s helium filling port without disturbing the rest of the setup.

\added{Since the cryostat employs an integrated He recondensing system, potential cryocooler-induced mechanical vibrations were considered as a limitation. To assess their influence, reference ODMR measurements were performed with and without the recondensing system in operation. Within the experimental resolution, no observable differences in ODMR linewidth, contrast, or overall spectral quality were detected. These observations indicate that possible cryocooler-induced vibrations do not significantly affect the long-distance ODMR measurements presented in this manuscript. However, such effects may become increasingly relevant for future implementations targeting diffraction-limited spatial imaging or scanning-based NV magnetometry, where sub-micron positional stability is required.}

\subsection{\label{sec:meth5}Measurement Electronics}
A continuous-wave 532\,nm laser with a typical output power of 100\,mW excites the NV centers at the bottom of the sample stick. The resulting red fluorescence is collected via a silicon avalanche photodiode. The resulting photovoltage that is proportional to the intensity of the fluorescence, is recorded using a National Instruments data acquisition card. Timing and synchronization are governed by a synchronous digital pattern and arbitrary waveform generator, which generates digital pulse sequences to modulate the microwave signal and control the timing of the analogue data acquisition. The sequence starts with a global trigger and proceeds with timing pulses synchronized to the microwave modulation as shown in Fig.\,\ref{fig:PulseSequenz} and explained further in the text.

\begin{figure}[t!]
    \centering
    \includegraphics[width=1\linewidth]{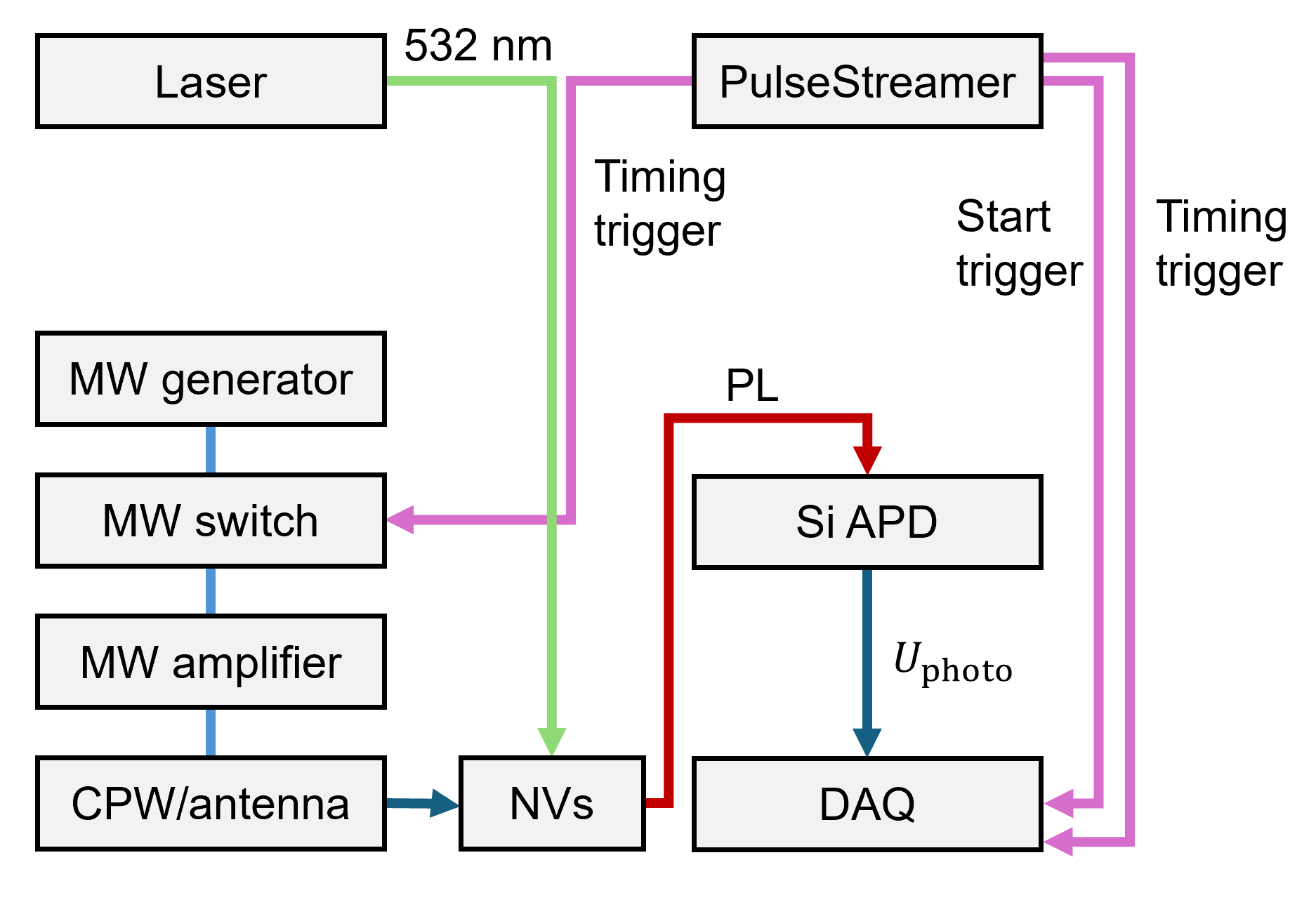}\\ 
\begin{tabular}{@{}l@{}l@{}}
\toprule
\parbox[t]{0.3\columnwidth}{\raggedright\textbf{Component}} & \parbox[t]{0.7\columnwidth}{\raggedright \textbf{Description}} \\
\midrule
\parbox[t]{0.3\columnwidth}{\raggedright Laser} & \parbox[t]{0.7\columnwidth}{\noindent\justifying Solid state laser Qioptiq Photonics, model NANO 250-532-100} \\
\parbox[t]{0.3\columnwidth}{\raggedright Pulse generator} & \parbox[t]{0.7\columnwidth}{\noindent\justifying Pulse Streamer 8/2, synchronous arbitrary waveform and digital pattern generator} \\
\parbox[t]{0.3\columnwidth}{\raggedright MW generator} & \parbox[t]{0.7\columnwidth}{\noindent\justifying Windfreak SynthHD v2, 10\,MHz–15\,GHz dual-channel microwave RF signal generator} \\
\parbox[t]{0.3\columnwidth}{\raggedright MW switch} & \parbox[t]{0.7\columnwidth}{\noindent\justifying Mini-Circuits, ZASWA-2-50DRA+} \\
\parbox[t]{0.3\columnwidth}{\raggedright MW amplifier} & \parbox[t]{0.7\columnwidth}{\noindent\justifying Mini-Circuits, high power amplifier HPA-25W-63+} \\
\parbox[t]{0.3\columnwidth}{\raggedright Si-APD} & \parbox[t]{0.7\columnwidth}{\noindent\justifying A-CUBE-S3000-10, $\varnothing$3\,mm, DC–10\,MHz} \\
\parbox[t]{0.3\columnwidth}{\raggedright DAQ module} & \parbox[t]{0.7\columnwidth}{\noindent\justifying National Instruments USB-6281} \\
\bottomrule
\end{tabular}

    \caption{Schematic diagram of the measurement electronics used for optically detected magnetic resonance (ODMR) measurements of the NV centers. The diagram illustrates the connections between components, with emphasis on the different trigger signals required for pulsed measurement mode, generated by a synchronous digital pattern and arbitrary waveform generator. Specific details on the devices may be found in the legend above.}
    \label{fig:MeasElectronics}
\end{figure}

\begin{figure}[t!]
    \centering
    \includegraphics[width=1\linewidth]{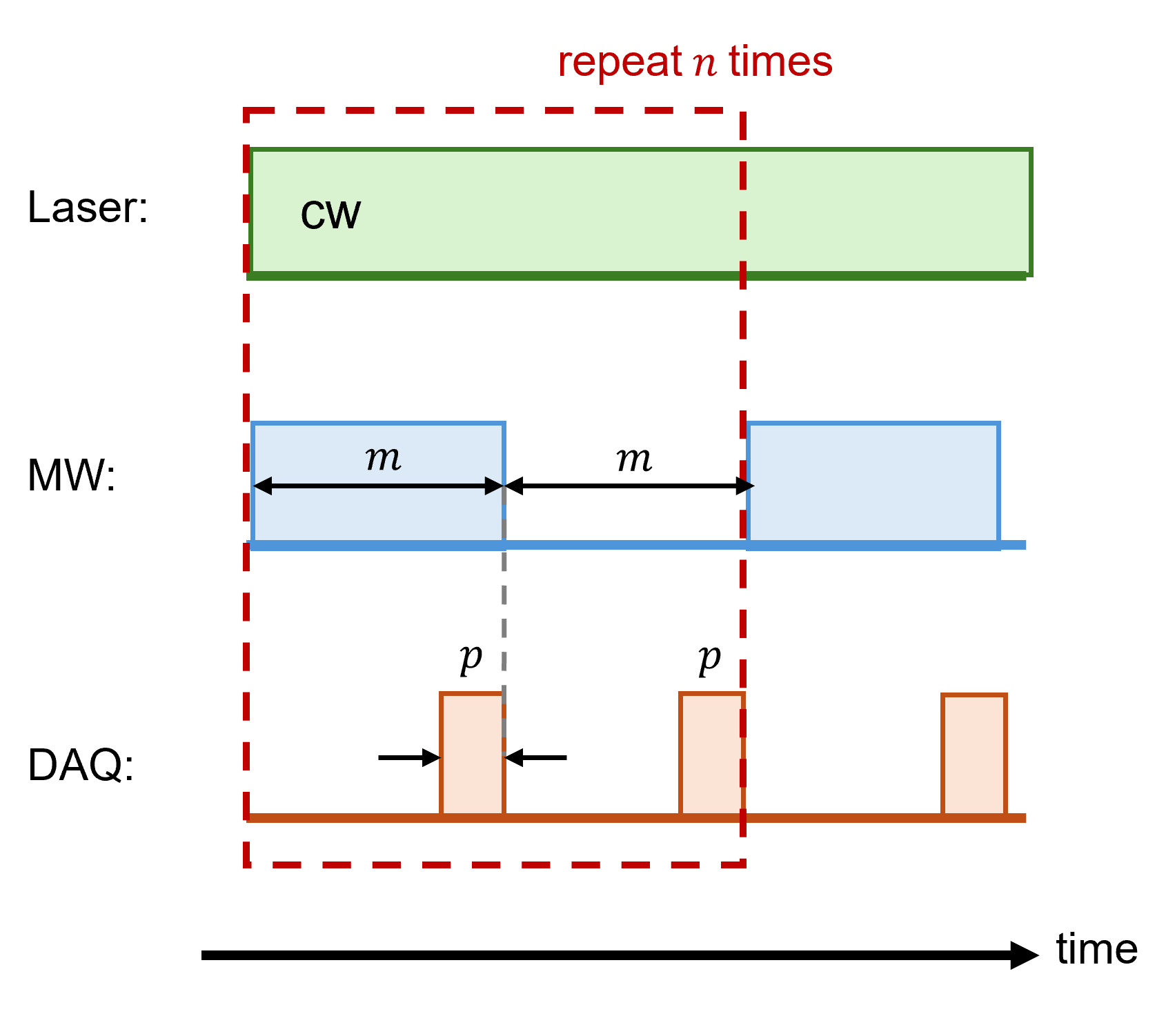}
    \caption{Pulsing sequence of the laser, microwave via the microwave switch, and data acquisition for ODMR measurements. The diagram shows the timing and synchronization of the laser pulses, microwave excitation, and data acquisition processes. The microwave is alternately switched on and off for $m=$16\,$\mu$s, while the detection frame is gated to $p=$1\,$\mu$s. This sequence is repeated $n=$ 10\,000 times for averaging. By pulsing the microwave in this way and evaluating the resulting values, only the spin-dependent fluorescence signal is isolated, while common-mode background and noise contributions are suppressed.}
    \label{fig:PulseSequenz}
\end{figure}

Microwave signals are produced by a signal generator, gated by a fast microwave switch synchronized to a digital pattern from an arbitrary waveform generator, and subsequently amplified. The signal is transmitted via coaxial cables to a coplanar waveguide (CPW) antenna on the sample stick, with losses along the path reducing the power delivered to the sample. The measurement electronics are schematically shown in Fig.\,\ref{fig:MeasElectronics} and the device details are listed in the table shown in the bottom part of Fig.\,\ref{fig:MeasElectronics}. The control of the measurement devices and data collection is governed by a LabVIEW program, which was chosen because our pre-existing cryostat control code was already implemented in LabVIEW, allowing seamless integration of hardware control and data acquisition.

In a typical measurement, the laser remains continuously on while the microwave excitation is pulsed. For each frequency step in the sweep, the analogue fluorescence signal is recorded both in the absence and presence of microwave driving, each lasting $m=$\,16\,$\mu$s. The detection is gated to a $p=$\,1\,$\mu$s window positioned within each interval, and the sequence is repeated $n=$\,10\,000 times for averaging, as illustrated in Fig.\,\ref{fig:PulseSequenz}. When the applied microwave frequency matches the spin resonance condition, transitions from the bright \( m_s = 0 \) state to the darker \( m_s = \pm1 \) states are driven, resulting in a drop in fluorescence intensity. The photovoltage signals recorded with and without microwave excitation are measured and summed separately. The point-by-point ratio of these signals as a function of microwave frequency yields the ODMR spectrum, from which the contrast is extracted through subsequent analysis and fitting. This normalization compensates for fluctuations in laser power or detector sensitivity and isolates the spin-dependent component of the fluorescence signal. The final data analysis is performed using a Python script, which processes and fits the resulting ODMR spectrum to extract the resonance frequency, magnetic field strength, linewidth, and contrast.

\section{\label{DataAnalysis}Data Analysis}

In order to demonstrate the performance and stability of our ODMR setup, we present typical data sets collected under various experimental conditions. To analyze the ODMR spectra, the resonances were fitted using two Lorentzian functions,

\begin{equation}
L(f) = A - \frac{C_{\pm}}{1 + \left( \frac{f - f_{\pm}}{\Gamma_{\pm}/2} \right)^2},
\end{equation}

\noindent where $A$ represents a constant background offset, $C_{\pm}$ the resonance depth, $f_{\pm}$ the resonance frequencies, and $\Gamma_{\pm}$ the full widths at half maximum (FWHM). The subscript $\pm$ indicates that a single resonance may split into two or more resonances due to a magnetic field, strain, or other perturbations. From these fits, the so called contrast $C_{\pm}$, linewidth $\Gamma_{\pm}$, and resonance frequencies $f_{+}$ and $f_{-}$ were extracted. These fit parameters provide a quantitative basis for comparing measurements across different conditions, such as varying temperature or magnetic field, as discussed in the following.

\begin{figure}[t!]
    \centering
    \includegraphics[width=1\linewidth]{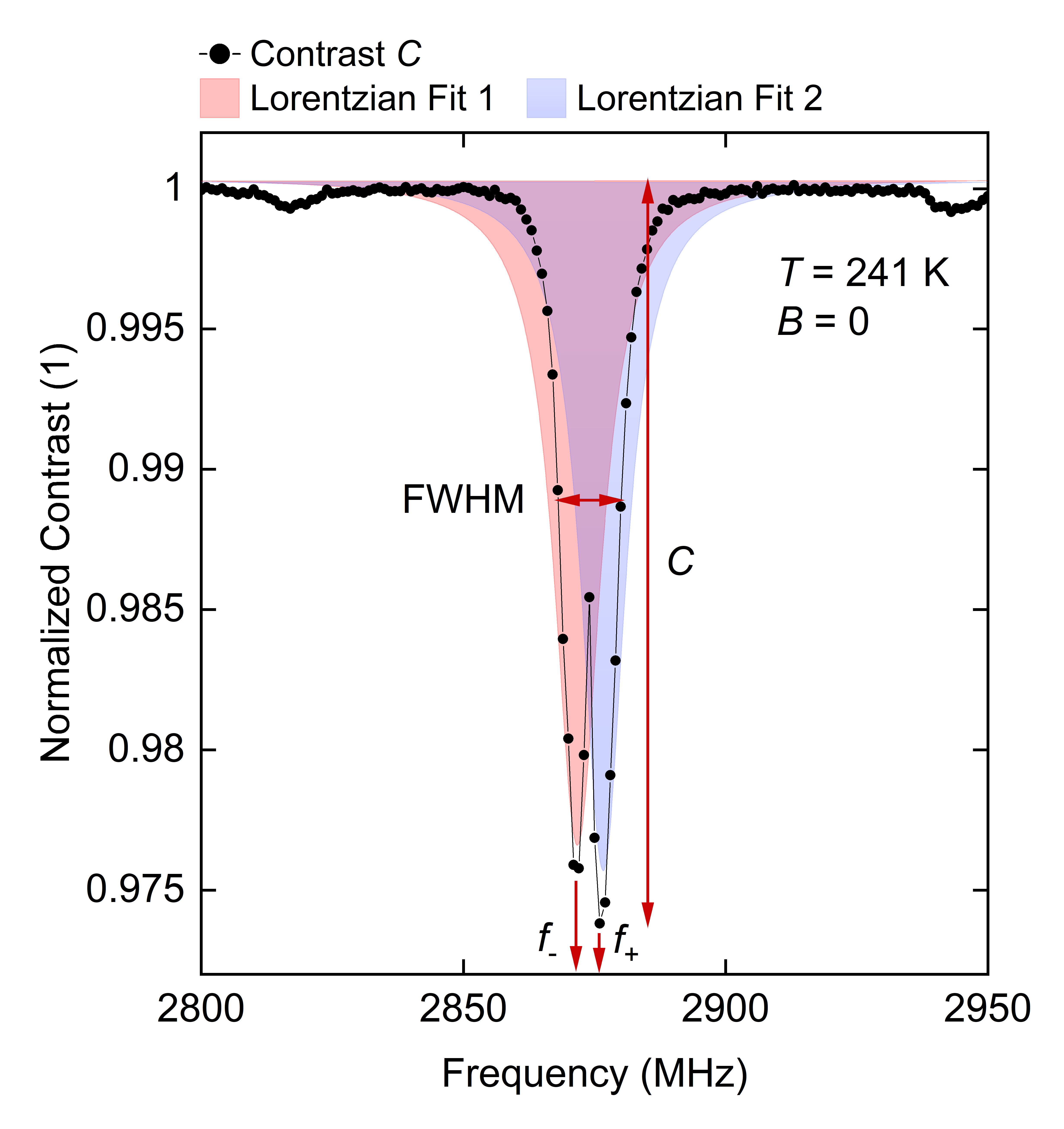}
    \caption{Typical ODMR spectrum of a bulk diamond chip with an NV concentration of 4.5\,ppm, measured at a temperature of 241\,K under nominal zero applied magnetic field. The spectrum was recorded with a microwave power of 4\,W and a laser power of 100\,mW. A split resonance is observed and fitted with two Lorentzian functions (blue and red). From the fit, we define the contrast, the full width at half maximum (FWHM), and the resonance frequencies $f_+$ and $f_-$. The apparent splitting at zero field is attributed to a combination of strain in the diamond lattice and a small remanent magnetic field from the superconducting 16\,T magnet of the cryostat.
}
    \label{fig:Figure75}
\end{figure}

In the ODMR spectrum shown in Fig.\,\ref{fig:Figure75}, recorded at 241\,K with 4\,W microwave \added{source} power and 100\,mW \added{source} laser excitation, a splitting of the resonance is observed even in nominal zero-field conditions. This apparent zero-field splitting may be attributed to two contributions. First, lattice strain in the diamond modifies the local crystal field around NV centers lifting the degeneracy of the $m_s = \pm1$ sublevels and resulting in a frequency separation of the resonances \citep{Batalov2009}. Second, a small \replaced{effective residual}{remanent} magnetic field \replaced{by}{from} the \deleted{surrounding} superconducting 16\,T magnet \replaced{may introduce}{introduces} an additional Zeeman splitting $\Delta\,f=f_+-f_-=4.37$\,MHz, \replaced{corresponding to an effective magnetic field}{This would correspond to a remanent magnetic field} of 0.078\,mT. \added{However, this field may be due to a combination of contributions, including Earth magnetic field ($\approx$ 0.0488\,mT in Munich~\citep{MagneticDeclinationMunich}), stray fields from surrounding experimental components, and remanent field of the superconducting magnet system.} Both \replaced{combine to}{effects manifest as} a visible separation of the resonance dips in the ODMR spectrum. Identifying these contributions early on is crucial as they determine the baseline for later analyses and underline the need to separate intrinsic material properties from external experimental influences.

\section{Temperature-dependent ODMR} \label{T_Dep}
\begin{figure}[t!]
    \centering
    \includegraphics[width=1\linewidth]{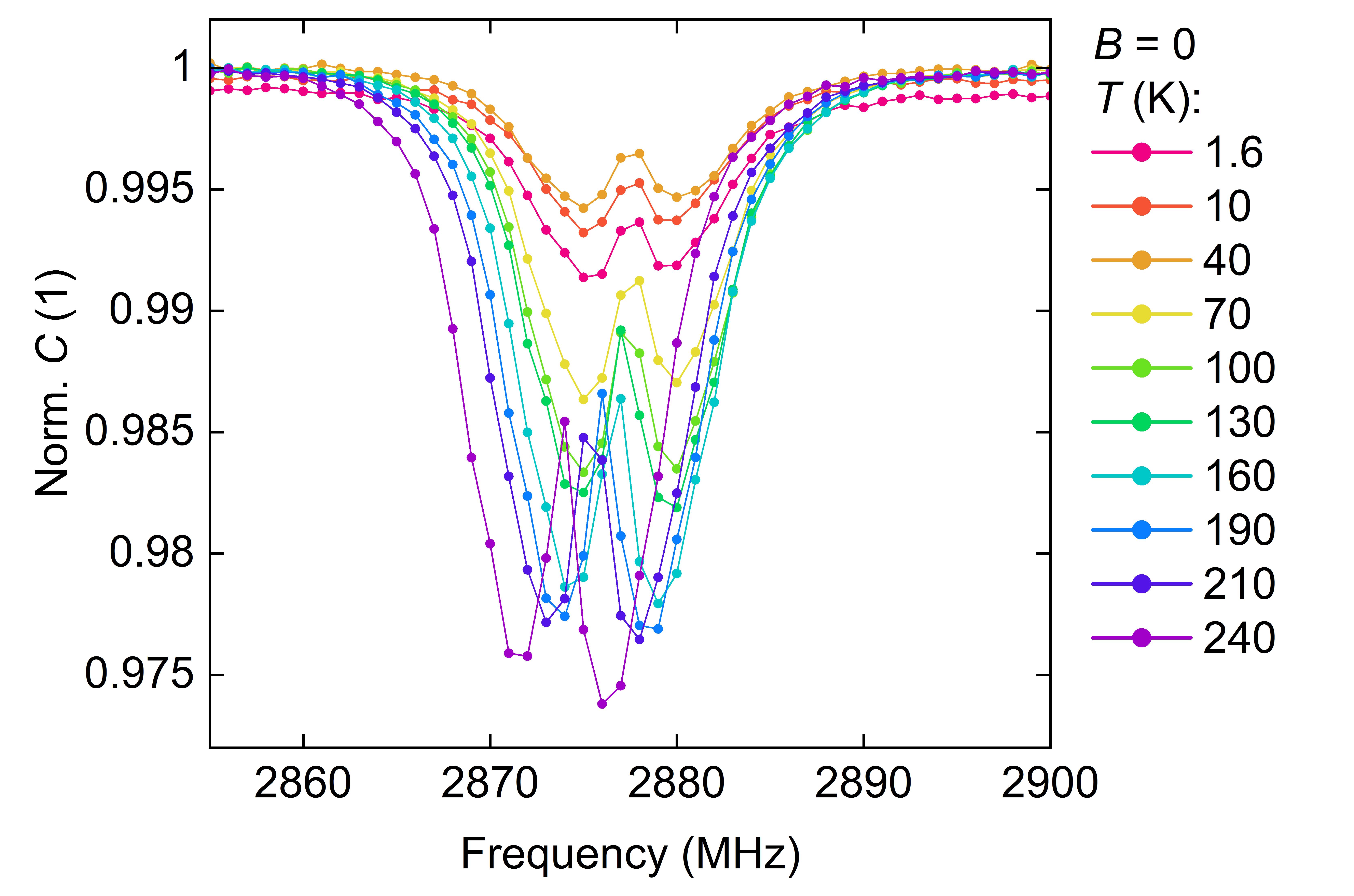}
    \caption{ODMR spectra of the bulk diamond chip (NV concentration 4.5\,ppm) measured at nominal zero applied magnetic field for different temperatures ranging from 1.6\,K to 240\,K. The measurements were performed with a microwave power of 4\,W and a laser power of 100\,mW. For clarity, only a subset of the spectra recorded is shown here, while the full data set is used in the analysis. The spectra exhibit a clear temperature dependence: the resonance contrast decreases with decreasing temperature, and the resonance center shifts to higher frequencies as the temperature decreases. As in Fig. \ref{fig:Figure75}, a splitting of the resonance is visible, originating from strain in the diamond lattice and the remanent magnetic field of the superconducting 16\,T magnet.}

    \label{fig:Figure55}
\end{figure}

\begin{figure}[htbp]
    \centering
    \includegraphics[width=1\linewidth]{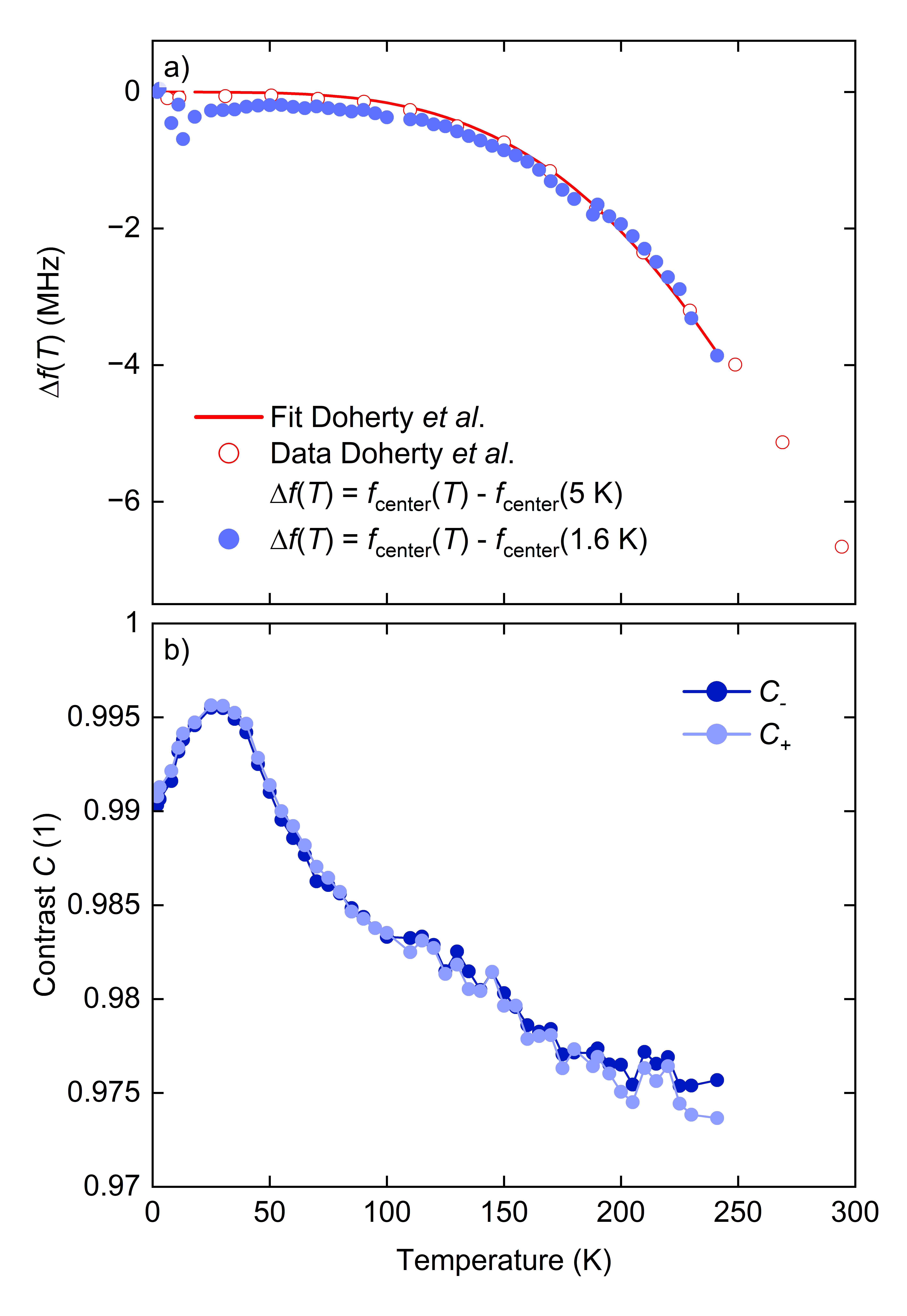}
    \caption{a) Temperature dependence of the ODMR resonance center.
    Frequency shift $\Delta f = f_{\text{center}}(T) - f_{\text{center}}(T_\text{ref})$ obtained from Lorentzian fits to the ODMR spectra. Our data (referenced to 1.6\,K) are shown together with the data and analytical fit from Doherty \textit{et al}. referenced to 5\,K) \citep{Doherty2014_Temp}, both showing a shift of the resonance center to lower frequencies with increasing temperature. b) ODMR contrast $C_{\pm}$ as a function of temperature. The contrast decreases with decreasing temperature, reaches a minimum near 30\,K, and slightly increases again at lower temperatures. This characteristic trend is consistent with the qualitative temperature dependence reported in the literature, including Ernst \textit{et al.}, and reflects the expected changes in NV spin-polarization and excited-state dynamics\,\citep{Ernst2023_TempContrast}.}
    \label{fig:Figure124}
\end{figure}

We recorded ODMR spectra of the diamond chip across a wide temperature range of 1.6–240\,K as shown in Fig.\,\ref{fig:Figure55}. For clarity, only a subset of the spectra recorded is shown here, while the full data set is used in the subsequent analysis. We observed that the resonance contrast decreases with decreasing temperature, while the resonance center shifts to lower frequencies. Across all temperatures, a residual splitting persists, arising from lattice strain and a small remanent field of the surrounding 16\,T magnet. Each spectrum was fitted with two Lorentzian functions and the midpoint between the resonances as well as the contrast $C_{\pm}$ were extracted. Following the procedure of Doherty \textit{et al}. \citep{Doherty2014_Temp}, the midpoint at the lowest temperature measured (1.6\,K in our case, 5\,K in their work) was subtracted from all data to obtain the relative shift $\Delta f(T)$. The resulting temperature dependence shown in Fig.\,\ref{fig:Figure124}\,a) together with measured data points and the analytical fit from \citep{Doherty2014_Temp}, agrees well with previous measurements and the analytical model of Doherty \textit{et al}. They showed that the shift of the NV ground state zero-field splitting parameter $\Delta\,f=f_{\text{center}}(T)-f_{\text{center}}(5\,\text{K})$ cannot be explained by thermal expansion alone. Instead, it arises predominantly from quadratic electron–phonon interactions, with thermal expansion providing only a minor contribution. 

\added{Good agreement is observed down to approximately 20\,K, below which deviations from the model become increasingly apparent. In this low-temperature region of laminar flow of the He gas, local heating due to laser illumination and microwave excitation may lead to a small temperature offset between the sample and the thermometer reading. To minimize such effects, the helium gas flow and microwave power were optimized during the measurements to reduce the local heating while maintaining a sufficient ODMR signal-to-noise ratio.}

As the temperature increases, phonons modulate the electronic orbitals of the NV center, thereby reducing the effective spin–spin interaction and shifting the resonance frequency.

Fig.\,\ref{fig:Figure124}\,b) shows the temperature dependence of the ODMR contrast of the bulk diamond chip, extracted from the spectra measured under nominal zero magnetic field in the VTI. The contrast decreases steadily with decreasing temperature, reaches a minimum around 30\,K, and then exhibits a slight increase at lower temperatures. This characteristic trend reflects the temperature dependence of NV spin-polarization efficiency and excited-state dynamics. At intermediate temperatures, thermally activated orbital and spin-mixing processes reduce the photoluminescence difference between the $m_s=0$ and $m_s=\pm1$ states, while at very low temperatures the suppression of phonon-mediated mixing leads to a partial recovery of contrast. A qualitatively similar behavior has been observed and analyzed in detail by Ernst \textit{et al.} in \citep{Ernst2023_TempContrast}. \replaced{These results reproduce the expected temperature-dependent behavior of NV centers over most of the investigated temperature range and demonstrate stable ODMR operation under cryogenic conditions, while small deviations at the lowest temperatures may arise from local heating effects that may be tracked, in principle, using an additional temperature sensor.}{These results reproduce the known temperature dependencies of NV centers and confirm the stable performance of the cryogenic setup.}

\section{Magnetic field-dependent ODMR} \label{Magn_Dep}
A superconducting magnet with a nominal maximum field of 16\,T, representative of a standard cryogenic magnet system, is used to generate the external magnetic field, with the sample positioned in the region of highest field homogeneity. \replaced{The magnet employs a bronze-routed multifilament Nb$_3$Sn configuration, in which Nb filaments are embedded in a CuSn (bronze) matrix, ensuring negligible flux jumps at low applied fields and correspondingly low remanent fields, which is particularly advantageous for sensitive optical measurements requiring reproducible field conditions~\citep{Wilson1983Cryogenics, Wilson1996Superconducting}.}{The magnet is wound with a bronze matrix, ensuring negligible flux pinning and correspondingly low remanent fields.} 

\replaced{In practice, the accessible magnetic field range for reliable ODMR measurements in the present setup is effectively limited to the mT regime. This limitation arises from two factors: first, the resonance frequencies shift linearly with increasing field, requiring microwave sources with increasingly large bandwidths. Second, the perpendicular field component inherent to the (100)-oriented diamond geometry induces spin-state mixing, which reduces the optical spin contrast and degrades the visibility and resolution of the ODMR resonance dips.}{In practice, however, the applied field is restricted to the mT range, as higher fields reduce the ODMR contrast and can impair the resolution of the resonance dips depending on the sample.} 

\begin{figure}[htbp!]
    \centering
    \includegraphics[width=1\linewidth]{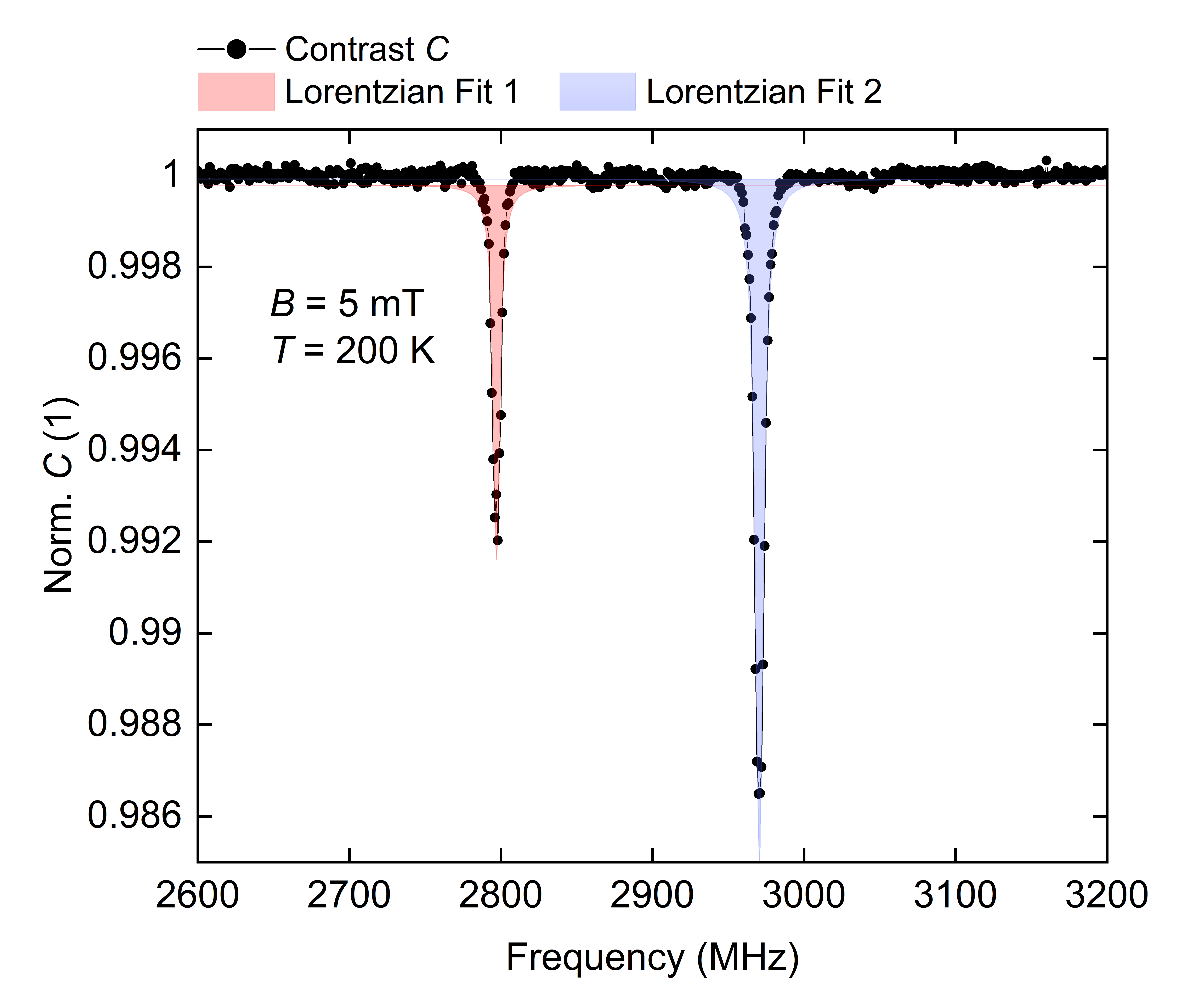}
    \caption{Typical ODMR spectrum measured at 200\,K under an applied magnetic field of 5\,mT. The resonances are fitted with two Lorentzian functions (red and blue), from which the full width at half maximum (FWHM), contrast, and resonance frequencies $f_+$ and $f_-$ are extracted. These quantities are further analyzed and discussed in the main text.}
    \label{fig:Figure84}
\end{figure}

A typical ODMR spectrum measured at 200\,K under an applied magnetic field of 5\,mT is shown in Fig.\,\ref{fig:Figure84}. As described earlier, the two resonances are fitted with Lorentzians (red and blue), yielding the resonance frequencies $f_+$ and $f_-$, the full width at half maximum (FWHM), and the contrast $C_{\pm}$. These parameters provide the basis for the quantitative analysis of field-dependent splitting, line broadening, and contrast reduction presented in the following.

\begin{figure}[hpbt!]
    \centering
    \includegraphics[width=1\linewidth]{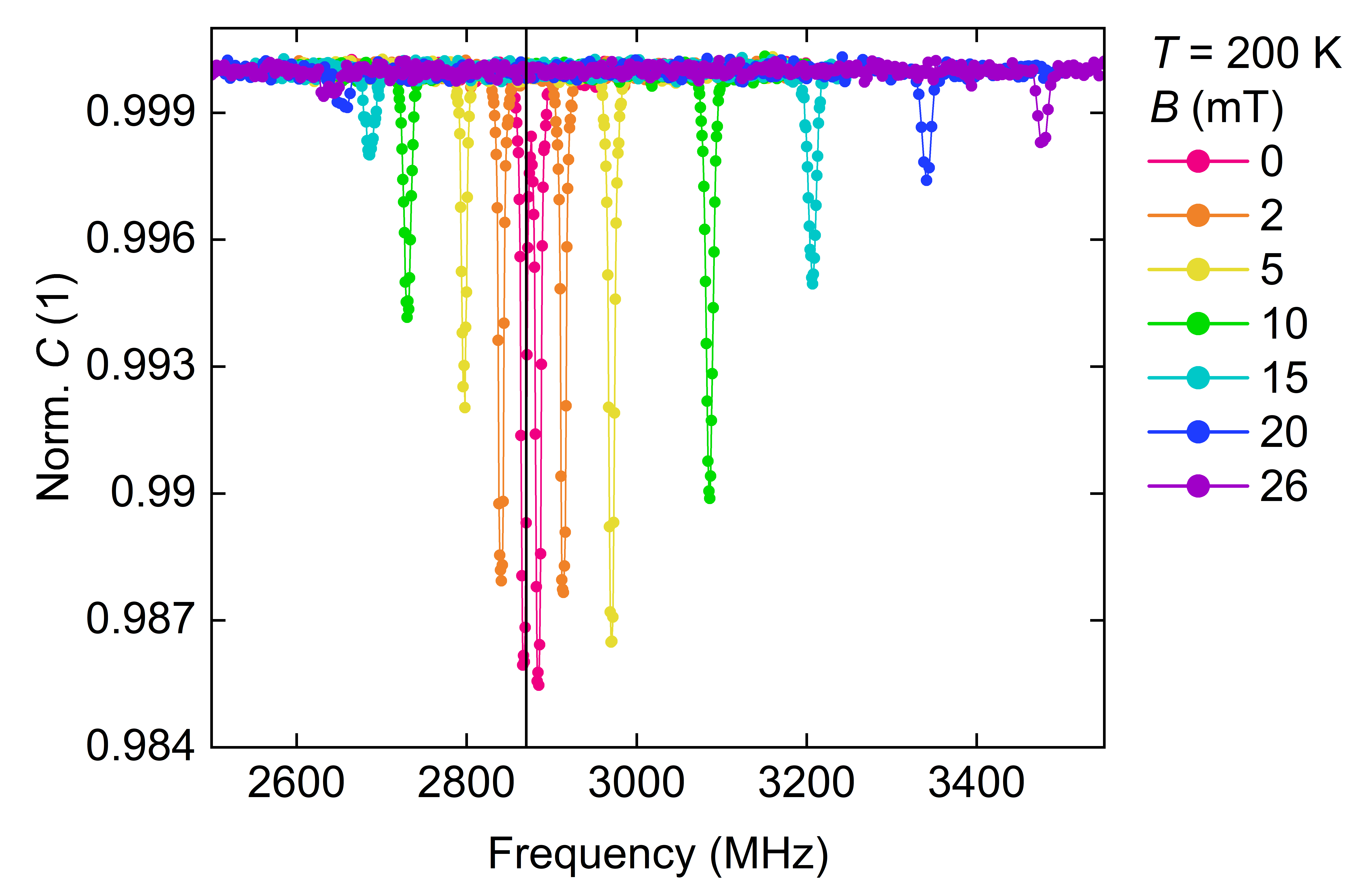}
    \caption{ODMR spectra of a high-density NV ensemble in bulk diamond recorded at 200\,K for different externally applied magnetic fields ranging from nominal zero field to 26\,mT. Microwave and laser powers were 4.0\,W and 100\,mW, respectively. A splitting of the resonance is visible even at zero field, which is attributed to strain in the diamond lattice and remanent magnetic fields from the superconducting magnet. With increasing external magnetic field, the splitting of the ODMR dips increases correspondingly. The resonance positions shift asymmetrically with field, likely due to perpendicular components of the magnetic field relative to the NV axes. A vertical black line marks the zero-field splitting at $D_\text{gs}=2.87$\,GHz as a reference for the unperturbed NV ground-state transition. The effects of strain and field orientation, and their implications for the interpretation of the $f_+$ and $f_-$ resonances, are discussed in detail in the main text and subsequent figures.}
    \label{fig:Figure44}
\end{figure}

In this section, we present a detailed analysis of magnetic-field-dependent ODMR spectra recorded at 200\,K on a bulk diamond chip with a (100) surface and external magnetic fields varied between 0 and 26\,mT oriented parallel to the (100) surface normal. Fig.\,\ref{fig:Figure44} shows the ODMR spectra for the NV ensemble. \replaced{As discussed in the previous section}{Even at nominal zero field}, a splitting of the resonance \added{at nominal zero field is observed.} \deleted{is evident, which we attribute to intrinsic strain in the diamond lattice as well as remanent magnetic fields from the sample magnet.} As the applied field increases, the splitting of the ODMR resonances grows correspondingly and the resonance positions shift asymmetrically, indicative of perpendicular components of the magnetic field relative to the NV axes as discussed in the following. The four NV orientations in the diamond chip with a (100)-oriented surface are symmetrically arranged at an angle of $\alpha=$ 54.7$^{\circ}$ to the surface normal. With the applied magnetic field along the normal [100], their resonances form two degenerate pairs producing the two distinct dips observed in the ODMR spectrum \citep{Zhang2018}.

\begin{figure}[hpbt!]
    \centering
    \includegraphics[width=1\linewidth]{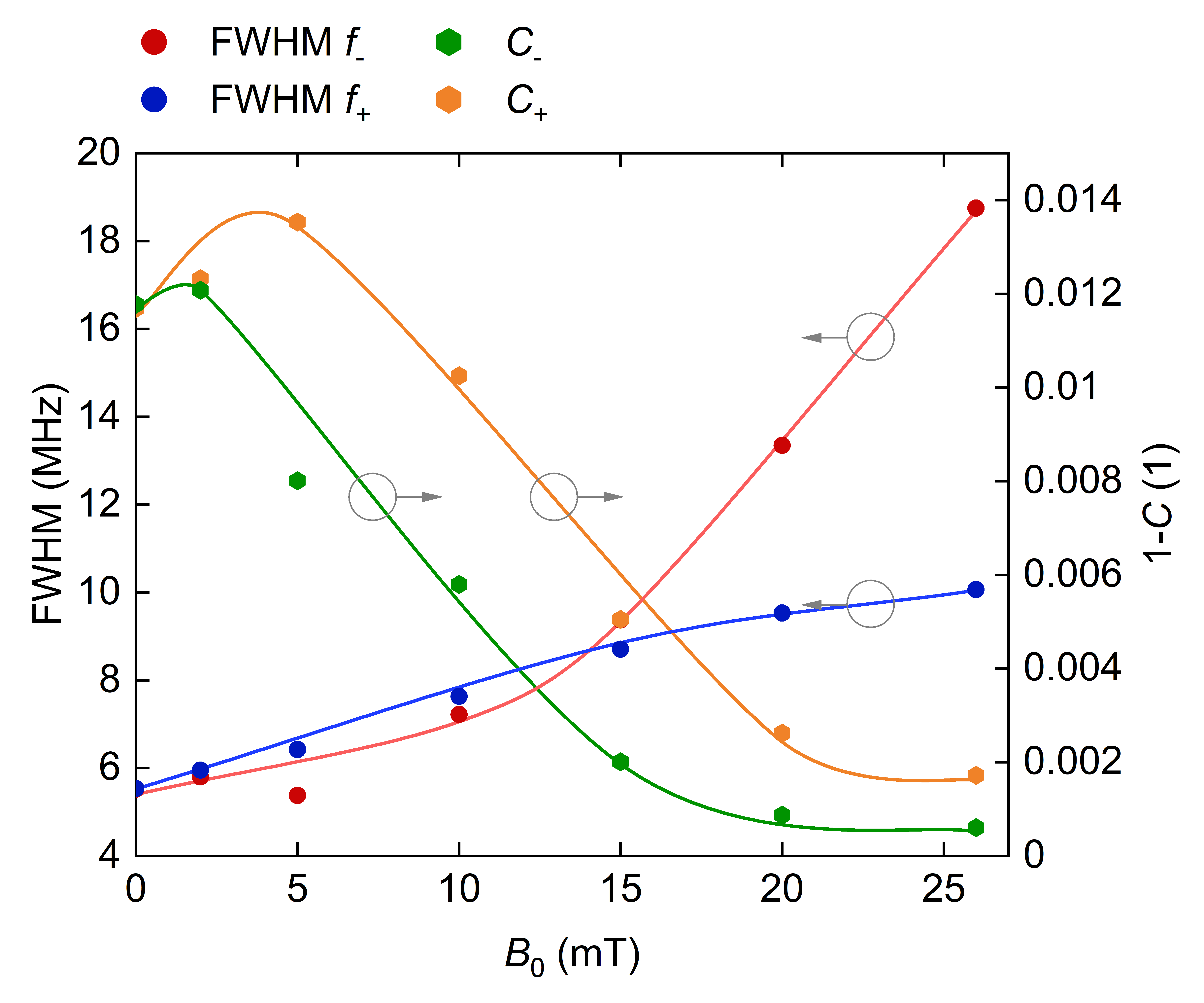}
    \caption{Full width at half maximum (FWHM) and inverted contrast ($1-C$) of the $f_+$ and $f_-$ resonances extracted from Lorentzian fits of the magnetic-field-dependent ODMR spectra. As the external magnetic field increases, the contrast of both $f_+$ and $f_-$ decreases, while the FWHM of the resonances increases. This behavior is attributed to increased inhomogeneous broadening in NV ensembles and partial misalignment of the NV axes relative to the applied magnetic field at higher fields. Lines serve to guide the eye.}
    \label{fig:Figure97}
\end{figure}

Next, we examine the evolution of the ODMR line shapes as summarized in Fig.\,\ref{fig:Figure97}. The full width at half maximum (FWHM) and inverted contrast $(1-C)$ of the $f_+$ and $f_-$ resonances reveal that increasing the external field causes the resonances to broaden while the contrast $C_{\pm}$ decreases. \replaced{This behavior can be attributed to a combination of increased inhomogeneous broadening within the NV ensemble and a reduction of effective ODMR contrast at higher magnetic fields. Inhomogeneous broadening arises from spatial variations in local strain, electric and magnetic fields, as well as crystal defects, leading to a distribution of resonance frequencies across different NV centers. In addition, for fields not perfectly aligned with a single NV axis, as is intrinsic to a (100)-oriented ensemble, the finite transverse field components induce magnetic-field-dependent mixing of spin states, which reduces the efficiency of optical spin polarization and spin-dependent fluorescence readout~\citep{Epstein2005DarkSpin, Horsthemke2024NVLifetime, Dolde_Thesis}. As a result, both the ODMR contrast decreases, and the apparent linewidth increases with increasing magnetic field. These effects emphasize that ensemble averaging over the four NV orientations and their distinct local environments must be carefully considered when interpreting high-field ODMR spectra.}{This behavior may be attributed to enhanced inhomogeneous broadening within the NV ensemble, which refers to the widening of the resonance caused by local variations in strain, electric and magnetic fields and crystal defects across different NV centers. In addition, partial misalignment of the NV axes relative to the applied field at higher fields contributes to the changes observed. These findings highlight that ensemble effects, which for NV centers arise from averaging over the four possible crystallographic orientations together with their distinct local strain and field environments, must be carefully considered when interpreting ODMR spectra.}

\begin{figure}[hpbt!]
    \centering
    \includegraphics[width=1\linewidth]{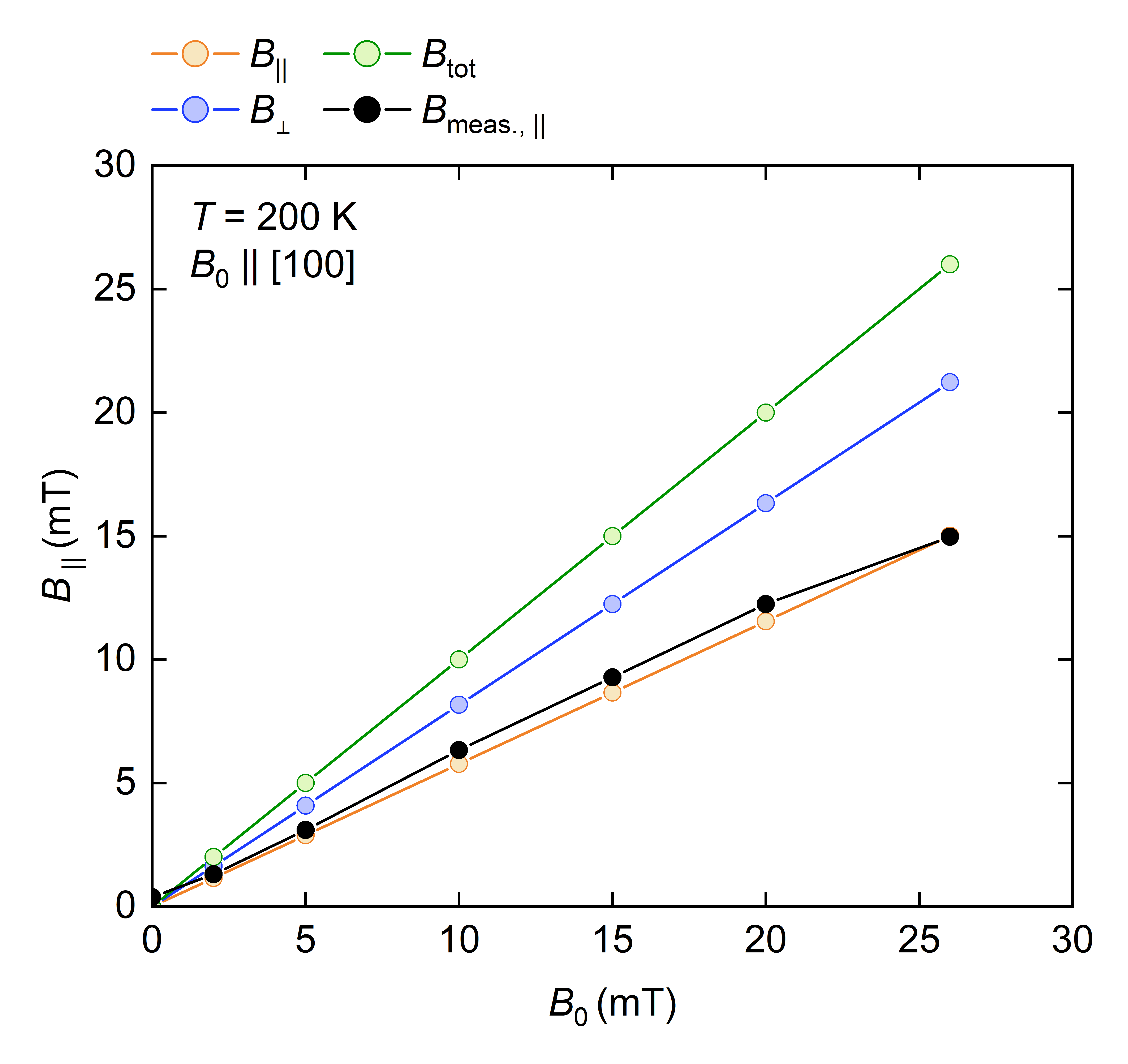}
    \caption{Calculated magnetic field components from the ODMR spectra of a bulk diamond chip with (100) surface. The parallel component $B_{\text{meas.},\,\parallel}$ is determined from the resonance splitting $\Delta\,f$ via $B_{\text{meas.},\,\parallel} = \Delta f / 2 \gamma$. For comparison, the expected parallel and perpendicular components, \added{$B_{\parallel}$ and $B_{\perp}$}, of an applied magnetic field $B_0$, oriented perpendicular to the diamond surface, are shown as $B_0 \cos(54.7^\circ)$ and $B_0 \sin(54.7^\circ)$, respectively, where the angle $\alpha = 54.7^\circ$ corresponds to the angle between the NV axis and the (100) diamond surface normal. The total field is calculated as $B_\mathrm{tot} = \sqrt{B_\mathrm{\parallel}^2 + B_\mathrm{\perp}^2}$. The experimentally derived parallel component from the ODMR splitting agrees well with the expected $B_0 \cos(54.7^\circ)$, confirming the expected NV orientation relative to the diamond surface.}
    \label{fig:Figure96}
\end{figure}

To quantify the field components, we extract the parallel and perpendicular contributions from the ODMR splitting as shown in Fig.\,\ref{fig:Figure96}. The measured parallel component $B_\mathrm{meas.,\,\parallel}$ is obtained via the frequency splitting $\Delta\,f$ as follows, $B_\mathrm{meas.,\,\parallel} = \Delta f / 2 \gamma$, with $\gamma=28$\,GHz\,T$^{-1}$ the gyromagnetic ratio. $B_\mathrm{meas.,\,\parallel}$ agrees well with the expected projection of the applied field, $B_0 \cos(54.7^\circ)$, confirming the anticipated NV orientation relative to the diamond surface. The total field, $B_\mathrm{tot} = \sqrt{B_\mathrm{theo.,\,\parallel}^2 + B_\mathrm{theo.,\,\perp}^2}$, further validates this agreement as it closely matches the magnitude of the applied magnetic field $B_0$, confirming that the parallel and perpendicular components extracted from the ODMR splitting $\Delta\,f$ correctly capture the vector projection of the field along and perpendicular to the NV axes. These results demonstrate that the ODMR splitting reliably reflects the geometric orientation of the NV centers under applied magnetic fields.

\added{The magnetic-field-dependent ODMR data are analysed starting from the NV ground-state spin Hamiltonian following Dolde et al.~\cite{Dolde2011} and Knauer et al.~\cite{Knauer2020}:}
 
\begin{align}
    H_\mathrm{gs} = \;&h\left(D_\mathrm{gs} + d^\parallel_\mathrm{gs}\sigma_z\right)
    \left[S_z^2 - \frac{1}{3}S(S+1)\right] 
    + \mu_B g_e \mathbf{S} \cdot \mathbf{B} \notag \\
    &- d^\perp_\mathrm{gs}\left[\sigma_x(S_x S_y + S_y S_x) 
    + \sigma_y(S_x^2 - S_y^2)\right]
    \label{eq:Ham}
\end{align}

\added{\noindent where $D_\mathrm{gs} = 2.87\,\mathrm{GHz}$ is the zero-field splitting, $d^\parallel_\mathrm{gs}$ and $d^\perp_\mathrm{gs}$ are the axial and non-axial components of the ground-state electric dipole moment, $\sigma$ is the local crystal strain field, which enters the Hamiltonian in the same mathematical form as an electric field, $\mu_B$ is the Bohr magneton, $g_e$ is the electron g-factor, and $S_x$, $S_y$, $S_z$ are the spin-1 operators. Since the axial dipole moment $d^\parallel_\mathrm{gs}$ is much smaller than the transverse component $d^\perp_\mathrm{gs}$, the axial strain term produces a negligible correction to $D_\mathrm{gs}$ and is dropped. Retaining only the transverse strain component perpendicular to the NV axis, parameterized as $E_\mathrm{strain}$ (in MHz) and entering through the $(S_x^2 - S_y^2)$ term, and splitting the magnetic field into components parallel and perpendicular to the NV axis, the Hamiltonian reduces to:}

\begin{equation}
    H_\mathrm{gs}=hD_\mathrm{gs}S_z^2+\mu_Bg_e \boldsymbol{B}\cdot \boldsymbol{S}+hE_\mathrm{strain}(S_x^2-S_y^2)
    \label{eq:HamMatrix}
\end{equation}

%

\added{\noindent This is the Hamiltonian from which the transition frequencies are derived.}
 
%
\added{Applying second-order perturbation theory to Eq.\,\eqref{eq:HamMatrix}, treating the perpendicular magnetic field component $B_\perp$ as the perturbation while retaining the full unperturbed energies including both $B_\parallel$ and $E_\mathrm{strain}$ in the energy denominators, the ODMR transition frequencies between $m_s = 0$ and $m_s = \pm 1$ are:}

\begin{align}
    f_\pm &= D_\mathrm{gs} \pm \sqrt{(\gamma B_\parallel)^2 
            + E_\mathrm{strain}^2} \notag \\
          &+ \frac{(\gamma B_\perp)^2}{2\left(D_\mathrm{gs} \pm 
            \sqrt{(\gamma B_\parallel)^2 + E_\mathrm{strain}^2}\right)} \notag \\
          &+ \frac{D_\mathrm{gs}\,(\gamma B_\perp)^2}
            {D_\mathrm{gs}^2 - (\gamma B_\parallel)^2 - E_\mathrm{strain}^2}
    \label{eq:freq}
\end{align}
 
\added{where $\gamma = 28\,\mathrm{GHz\,T^{-1}}$ is the NV gyromagnetic ratio, and $B_\parallel$ and $B_\perp$ are the magnetic field components parallel and perpendicular to the NV symmetry axis, respectively. For a (100)-oriented diamond with the field applied along the surface normal, all four NV families are inclined at $54.7^\circ$ with respect to the field direction, giving $B_\parallel = B_0\cos(54.7^\circ) \approx 0.577\,B_0$ and $B_\perp = B_0\sin(54.7^\circ) \approx 0.816\,B_0$.}
 
\added{The first and second term of Eq.\,\eqref{eq:freq} describes the symmetric Zeeman and strain contributions, which shift $f_+$ and $f_-$ equally in opposite directions and leave the midpoint fixed at $D_\mathrm{gs}$. The third and fourth terms are the second-order corrections due to $B_\perp$: since the denominators differ between $f_+$ and $f_-$, these corrections are unequal for the two transitions, causing both resonances to shift by different amounts and in the same net direction. The midpoint between $f_+$ and $f_-$ therefore shifts with increasing field, which is the physical origin of the asymmetric behaviour observed experimentally.}
 
\begin{figure}[hpbt!]
    \centering
    \includegraphics[width=1\linewidth]{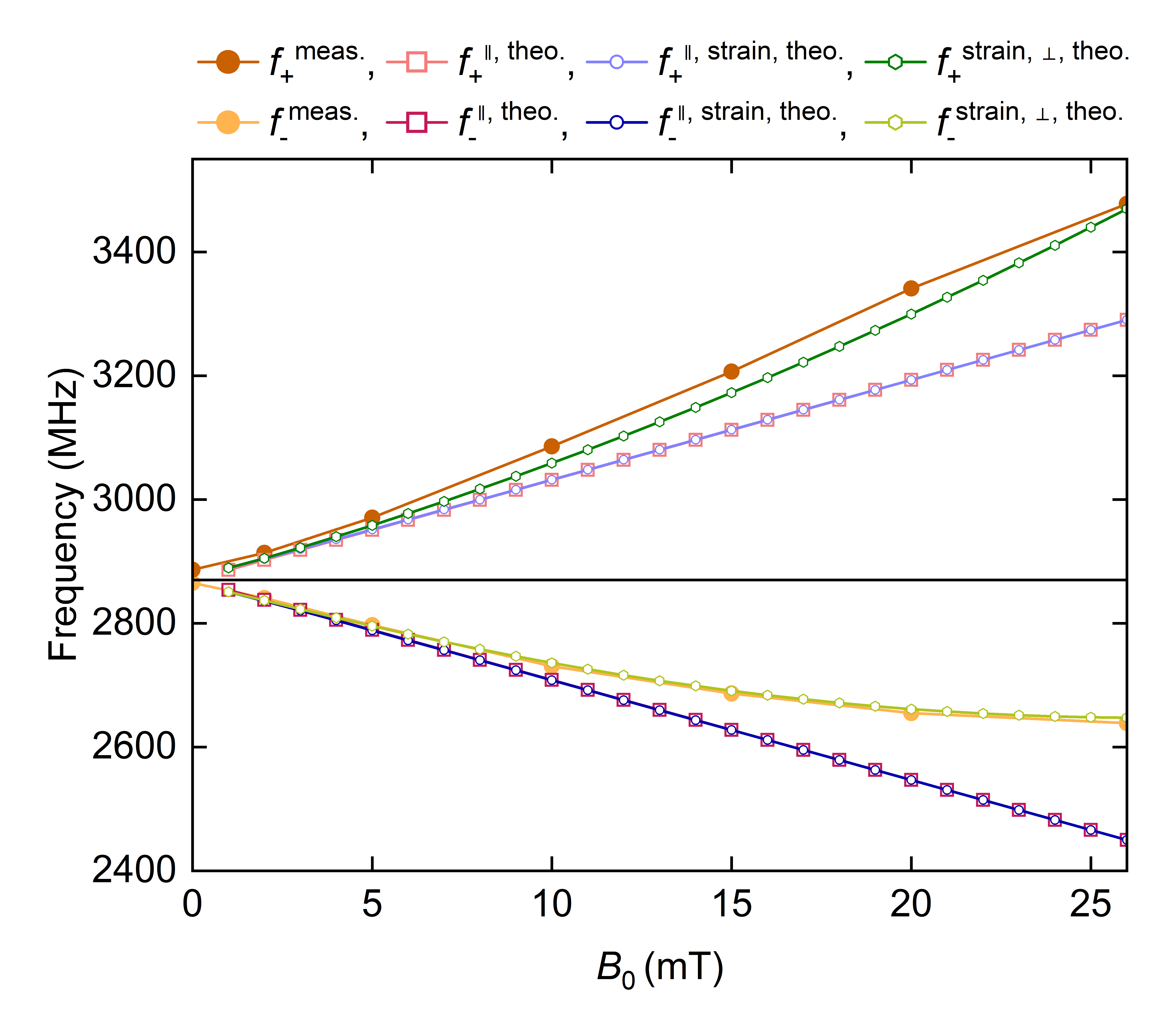}
    \caption{ODMR resonance frequencies $f_\pm$ as a function of the applied magnetic field $B_0$ oriented perpendicular to the diamond surface (100) normal. The expected frequencies are shown for three theoretical considerations: (i) considering only the orientation of the NV axes relative to the (100) diamond surface normal, $f_\pm^{\text{$\parallel$,\,theo.}}$, (ii) including the zero-field splitting due to strain, $f_\pm^{\text{$\parallel$,\,strain,\,theo.}}$, and (iii) including both strain and the perpendicular component of the magnetic field using second-order perturbation theory, $f_{\pm}^{\text{strain},\,\perp,\,\text{theo.}}$. The experimentally extracted resonance frequencies $f_\pm^{\text{meas.}}$ from the ODMR spectra are plotted for comparison. With increasing magnetic field, the measured data show a clear increase in the frequency splitting between $f_+$ and $f_-$. The calculated curves for the purely orientation-based and strain-corrected cases overlap closely, indicating that strain contributes only weakly to the overall frequency shift. In contrast, including the perpendicular component of the magnetic field introduces an asymmetric shift of the resonance frequencies, reproducing the qualitative behavior observed experimentally. Remaining deviations are attributed to ensemble effects arising from the superposition of signals from all NV orientations.
}
    \label{fig:Figure89}
\end{figure}

\added{Three scenarios are considered in Fig.~\ref{fig:Figure89} in the comparison with the measured frequencies $f_{\pm}^{\text{meas.}}$: (i) the simple geometric model accounting only for the NV axis orientation relative to the (100) surface normal, $f_\pm^{\parallel,\,\text{theo.}}$; (ii) the inclusion of a strain-induced zero-field splitting, 
$f_\pm^{\parallel,\,\text{strain,\,theo.}}$; and (iii) the full expression Eq.\,\eqref{eq:freq} including strain and the second-order 
perpendicular-field contribution for a single NV orientation, $f_{\pm}^{\text{strain},\,\perp,\,\text{theo.}}$. The strain parameter $E_\mathrm{strain}$ is treated as a free parameter, as the present measurements are performed on an NV ensemble rather than a single NV centre, and an independent determination of the strain is beyond the scope of this work.}
 
\added{The full expression Eq.\,\eqref{eq:freq} provides an excellent description of the experimentally observed frequency shifts, reproducing both the asymmetric displacement of the midpoint and the overall field dependence of both resonances.}

\deleted{Finally, Fig.14 compares the experimentally extracted resonance frequencies $f_{+}$ and $f_{-}$ with theoretical predictions. Building on the established theoretical framework for strain- and perpendicular-field–induced corrections to the NV ground-state splitting\,(Dolde2011, Knauer2020), it becomes essential to include a second-order correction to the simple Zeeman model. In particular, the perpendicular field component $B_{\perp}$ induces an additional resonance shift proportional to $(\gamma B_{\perp})^2 / D_{\mathrm{gs}}$, with $D_\mathrm{gs} = 2.87\,\mathrm{GHz}$ being the zero-field splitting of the ground state.}

\deleted{Three scenarios are considered in the comparison: (i) the NV axis orientation relative to the (100) surface normal, $f_\pm^{\text{(100),\,theo.}}$ , (ii) the inclusion of strain-induced zero-field splitting, $f_\pm^{\text{strain,\,theo.}}$, and (iii) the combined effect of strain and the second-order perpendicular-field contribution for a single NV orientation, $f_{\pm}^{\text{strain},\,\perp, \text{theo.}}$.}

\deleted{For clarity, Fig.15 displays only the latter model alongside the experimentally obtained frequencies.We note that this constitutes a simplified representation of the second-order perturbation calculation providing a qualitative description of the asymmetric contribution observed in the experimental data, while not capturing all ensemble or higher-order effects.The experimental data display a pronounced asymmetry in the midpoint between $f_+$ and $f_-$ which is qualitatively reproduced by the second-order perturbation calculation for a single NV axis. The remaining deviations between experiment and theory are likely due to ensemble averaging over multiple NV orientations. Overall, these results highlight the importance of accounting for both the geometric orientation of NV centers and strain when interpreting magnetic-field-dependent ODMR spectra in diamond.}

\section{ODMR Measurements on Strontium Ruthenate}\label{SampleSRO}
SrRuO$_3$ is a $4\,d$ transition-metal oxide crystallizing in the orthorhombic perovskite structure with space group Pbnm\,\citep{BENSCH1990171}. It is a ruthenate that exhibits ferromagnetic order at ambient pressure conditions, with a Curie temperature of approximately 165\,K. This well-defined magnetic transition, together with the material’s strong uniaxial anisotropy and comparatively simple domain structure\,\citep{Kunkemöller_PhysRevB.96.220406}, makes SrRuO$_3$ a good benchmark system for evaluating the performance of our NV-based magnetometry setup under cryogenic conditions relevant for quantum materials research.

\begin{figure}[hpbt!]
    \centering
    \includegraphics[width=1\linewidth]{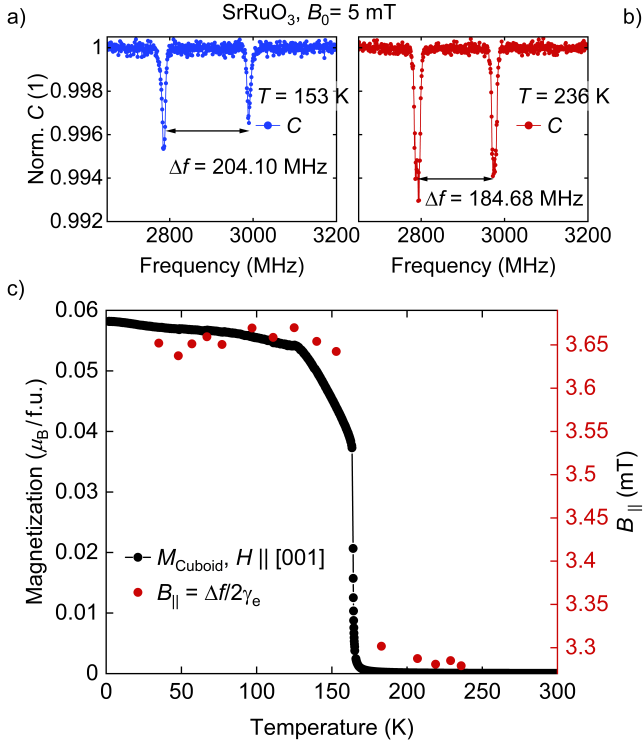}
    \caption{Comparison of SrRuO$_3$ magnetization measured in a Quantum Design MPMS\,3 and the parallel component of the magnetic field inferred from the NV resonance splitting in a diamond with a (100) surface. (a, b) ODMR spectra recorded above (238\,K) and below (153\,K) the magnetic transition temperature under a 5\,mT bias field, using 100\,mW laser power and 5\,W MW power. The frequency difference $\Delta\,f$ between the Zeeman-split resonances is indicated. (c) Magnetization data measured on a cuboid SrRuO$_3$ sample with the magnetic field applied parallel to the [001] crystallographic direction under a 5\,mT field-cooled bias corresponding to the bias field used in the ODMR measurements. Both data sets reveal the magnetic transition at $T_C \approx 164$\,K, the MPMS data show a sudden rise in magnetization, while the NV resonance splitting exhibits an increase near the same temperature.}
    \label{fig:Figure93}
\end{figure}

Fig.\,\ref{fig:Figure93} illustrates the performance of our NV-based ODMR setup in detecting the magnetic response of ferromagnetic SrRuO$_3$. \added{The investigated SrRuO$_3$ sample had dimensions of approximately 1.7\,mm $\times$ 0.68\,mm $\times$ 0.16\,mm and was positioned between the microwave stripline and the diamond chip. The crystallographic [110]-direction of the sample was aligned parallel to the stripline, while the [001]-direction was oriented along the out-of-plane $z$-component of the applied magnetic field.} Panels\,(a) and (b) show representative ODMR spectra recorded at 153\,K and 236\,K, below and above the magnetic transition temperature $T_C = 164$\,K, under a 5\,mT bias field. A clear change in the splitting of the resonance is observed across the transition. Panel\,(c) summarizes the extracted splittings as a function of temperature. These are shown together with magnetization data recorded in a Quantum Design MPMS\,3 on a cuboid sample with the field applied along the pseudo-cubic [001] direction under the same 5\,mT field-cooled bias. The magnetization exhibits the ferromagnetic transition at $T_C = 164$\,K, visible as a sharp increase upon cooling and subsequent saturation at lower temperatures. The emergence of an additional splitting in the ODMR spectra below this temperature indicates an additional magnetic field contribution, which we attribute to the onset of ferromagnetic order. These observations confirm that our setup reliably detects bulk magnetic transitions via the local magnetic field sensed by the NV ensemble and validate the sensitivity and robustness of our NV magnetometry platform for probing temperature-dependent magnetic phenomena in correlated materials.

\section{Sensitivity of ODMR-based NV Magnetometry}\label{SensitivityEstimate}
Our setup demonstrates reliable performance under demanding cryogenic and optical conditions, as evidenced by its ability to reproduce the well-established temperature- and field-dependent spin-resonance behavior of NV centers in diamond as well as the ferromagnetic transition of SrRuO$_3$. These benchmark measurements validate both the technical stability and the scientific capability of the system, establishing a strong foundation for future investigations of more complex materials and more challenging experimental environments. In particular, NV-based magnetometry has the potential to complement conventional \added{bulk magnetometry techniques such as VSM and} SQUID measurements by providing nanoscale spatial resolution and sensitivity to local magnetic noise, enabling the detection of fluctuations and inhomogeneities that are typically averaged out in \added{bulk} SQUID experiments. 

\added{With a $T_2^* = 0.5\,\mu$s for our commercial diamond chip, we estimate a lower bound on the field sensitivity of ${\sim}230\,\text{nT}/\sqrt{\text{Hz}}$ under moderate experimental conditions, corresponding to a minimum resolvable field of 
${\sim}230\,\text{nT}$ at a measurement time of $\tau = 1\,\text{s}$, and detectable magnetic moments of ${\sim}10^{-22}$--$10^{-21}\,\text{Am}^2$ at NV-to-sample distances of 50--100\,nm, which is an improvement of 10--11 orders of magnitude 
over VSM (sensitivity ${\sim}10^{-11}\,\text{Am}^2$\,\citep{QuantumDesignMPMS3}). As a concrete example, FIB-prepared SrRuO$_3$ lamellae with typical dimensions of 
$20 \times 10 \times 0.1\,\mu\text{m}^3$ carry an magnetic moment of ${\sim}3 \times 10^{-12}\,\text{Am}^2$ at low temperatures, which lies approximately a factor of 25 below the VSM detection limit yet remains well within the detection range of our ODMR setup --- at a sensor distance of 100\,nm the lamella produces an estimated local stray field of ${\sim}\,50-100\,\text{mT}$, exceeding the ODMR detection threshold by more than five orders of magnitude. This sensitivity advantage arises not from superior field resolution in absolute terms, but from the nanometre proximity of the NV sensor to the sample, where the local stray field of even a nanoscale specimen far exceeds the detection threshold --- a regime fundamentally inaccessible to macroscopic induction-based techniques. This highlights the potential of NV-ensemble magnetometry as a sensitive local probe for nanostructured samples and geometrically confined systems, including the investigation of interface and surface magnetism in correlated oxide thin films. A detailed sensitivity analysis will be provided elsewhere~\citep{AnhTong2026Thesis}.}

\section{Conclusions}\label{Discussion_Concl}
We have presented a modular ODMR setup that overcomes the intrinsic challenges of implementing optical experiments within standard helium bath cryostats featuring variable temperature inserts. The setup achieves efficient optical coupling over an extended free-space path length approaching two meters from the optical head to the sample, a significant technical hurdle given the spatial constraints and mechanical instabilities inherent to cryogenic environments.

Central to the design are three key components: (1) the bespoke sample stick, carefully engineered to fit within the 30\,mm diameter VTI bore while integrating both microwave delivery and optical components, (2) the rail-guided platform with spring-loaded vertical fine adjustment and precise lateral positioning ensuring reproducible alignment and mitigating mechanical stress during thermal cycling, and (3) the support frames that enable pre-alignment and testing at room temperature significantly reducing experimental downtime.

This modular design not only facilitates stable and reproducible NV-based ODMR measurements across a wide temperature range of 1.6\,K to 300\,K but also offers flexibility for integration with alternative cryogenic platforms or more complex sample environments. \added{While reliable temperature control is achieved down to approximately 20\,K, the onset of laminar helium gas flow in this temperature regime requires careful optimization of the microwave power and He gas flow conditions in order to minimize local heating effects at the sample position.} In particular, the mechanical and optical architecture is well-suited for adaptation to diamond anvil cells where space constraints and the need for stable optical access are even more severe. Future efforts will aim to further enhance the setup, optimizing microwave delivery and optical coupling to enable robust and high-fidelity NV-based measurements in diamond anvil cells.

By documenting detailed assembly and alignment procedures, we aim to provide a practical guide for researchers seeking to implement similar setups in existing cryostats, reducing barriers to entry for cryogenic NV experiments.

Overall, our work bridges a crucial gap between conventional cryogenic infrastructure and advanced optical measurement techniques, enabling new investigations of magnetic materials and phenomena under controlled low-temperature conditions with minimal compromise on optical performance.

\begin{acknowledgments}
We wish to acknowledge the support and valuable contributions of C. Lüthi, J. Finley, F. Reinhard, M. Brandt, L. Todenhagen, L. Hanschke, C. Zu, G. He, M. Lampl, S. Giemsa, \added{A. Hahn,} and S. Mayr. We gratefully acknowledge M. Braden, K. Jenni, and S. Kunkemöller for providing the SrRuO$_3$ sample used in this study. We are also grateful to the mechanical workshop of the Technical University of Munich for their expert assistance in the fabrication of \replaced{bespoke}{custom} components essential to this work. This study received funding from the Deutsche Forschungsgemeinschaft (DFG, German Research Foundation) under TRR80 (From Electronic Correlations to Functionality, Project No. 107745057), TRR360 (Constrained Quantum Matter, Project No. 492547816, Projects A6, C1, and C3), SPP2137 (Skyrmionics, Project No. 403191981, Grant PF393/19), and the excellence cluster MCQST under Germany’s Excellence Strategy EXC-2111 (Project No. 390814868). Financial support by the European Research Council (ERC) through Advanced Grants No. 291079 (TOPFIT) and No. 788031 (ExQuiSid) is gratefully acknowledged.
\end{acknowledgments}

\section*{Conflict of Interest Statement}
The authors have no conflicts to disclose.

\section*{Data Availability Statement}
Data will be made available on reasonable request.

\section*{CRediT authorship contribution statement}
\textbf{AT}: Writing - original draft, Conceptualization, Data Curation, Formal Analysis, Investigation, Methodology, Software, Visualization, \textbf{AB}: Review \& Editing, \textbf{MK}: Software, \textbf{JS}: Resources, Review \& Editing, \textbf{CB}: Resources, Review \& Editing \textbf{KB}: Resources, Review \& Editing \textbf{FF}: Resources, \textbf{DB}: Resources, Validation, Review \& Editing \textbf{CP}: \added{Project Proposal}, Supervision, Conceptualization, Validation, Review \& Editing.
\newpage
\bibliography{bibliography}

\end{document}